\documentclass[11pt]{article}
\usepackage[a4paper, margin=2.5cm]{geometry}
\date{}

\usepackage{fvextra}
\usepackage[T1]{fontenc}
\usepackage[normalem]{ulem} 

\usepackage[usenames,dvipsnames,svgnames,x11names]{xcolor}

\newcommand{\pgs}[1]{{\textcolor{red}{~~\em #1~pgs}}}
\renewcommand{\pgs}[1]{}

\usepackage[nocompress]{cite}
\usepackage{balance}
\usepackage{listings}

\lstdefinelanguage{XML}
{
basicstyle=\ttfamily\footnotesize,
  morestring=[b]'',
  moredelim=[s][\bfseries\color{Maroon}]{<}{\ },
  moredelim=[s][\bfseries\color{Maroon}]{</}{>},
  moredelim=[l][\bfseries\color{Maroon}]{/>},
  moredelim=[l][\bfseries\color{Maroon}]{>},
  morecomment=[s]{<?}{?>},
  morecomment=[s]{<!--}{-->},
  commentstyle=\color{gray},
  stringstyle=\color{blue},
  identifierstyle=\color{red},
}

\usepackage{moreverb}
\usepackage[nounderscore]{syntax}
\usepackage[pdftex]{graphicx}
\usepackage{subfig}
\usepackage{wrapfig}
\graphicspath{{./figures/}}
\DeclareGraphicsExtensions{.pdf}
\usepackage[cmex10]{amsmath}
\usepackage{amssymb}
\usepackage{mathtools}
\usepackage{amsthm}
\usepackage{amsfonts}
\usepackage{gensymb}

\usepackage{subfig}
\usepackage{algorithmicx}
\usepackage{algpseudocode}
\usepackage[ruled]{algorithm}

\definecolor{light-gray}{gray}{0.75}
\algrenewcommand{\algorithmiccomment}[1]{\hskip3em{{\footnotesize \textcolor{light-gray}{$\blacktriangleright$}}} #1}

\usepackage{multirow} 
\usepackage{rotating}
\usepackage{booktabs}
\usepackage{colortbl}
\usepackage{tablefootnote} 
\usepackage{array}

\usepackage[pdftex,colorlinks=true,urlcolor=blue,citecolor=blue]{hyperref}
\usepackage{xspace}
\usepackage{blindtext}

\usepackage{enumitem}
\newcommand{\agentx}{AgentX\xspace}

\begin{document}

\title{AgentX: Towards Orchestrating Robust Agentic Workflow Patterns with FaaS-hosted MCP Services}

\author{
Shiva Sai Krishna Anand Tokal$^{1}$,
Vaibhav Jha$^{1}$,\\
Anand Eswaran$^{2}$,
Praveen Jayachandran$^{2}$ and
Yogesh Simmhan$^{1}$\\
\small 
$^{1}$~\textit{Indian Institute of Science, Bangalore}\\
\small 
$^{2}$~\textit{IBM India Research Lab, Bangalore}\\
\small 
Email: \{shivatokal, vaibhavjha, simmhan\}@iisc.ac.in
}

\maketitle

\begin{abstract}
\textit{Generative Artificial Intelligence (GenAI)} has rapidly transformed various fields including code generation, text summarization, image generation and so on. Agentic AI is a recent evolution that further advances this by coupling the decision making and generative capabilities of LLMs with actions that can be performed using tools. While seemingly powerful, Agentic systems often struggle when faced with numerous tools, complex multi-step tasks, and long-context management to track history and avoid hallucinations. Workflow patterns such as Chain-of-Thought (CoT) and ReAct help address this. Here, we define a novel agentic workflow pattern, \textit{\agentx}, composed of stage designer, planner, and executor agents that is competitive or better than the state-of-the-art agentic patterns. We also leverage Model Context Protocol (MCP) tools, and propose two alternative approaches for deploying MCP servers as cloud Functions as a Service (FaaS). We empirically evaluate the success rate, latency and cost for \agentx and two contemporary agentic patterns, ReAct and Magentic One, using these the FaaS and local MCP server alternatives for three practical applications. This highlights the opportunities and challenges of designing and deploying agentic workflows.
\end{abstract}

\section{Introduction\pgs{2.5}}
Generative Artificial Intelligence (GenAI)~\cite{genai1,genai2} has rapidly transformed various fields including code generation, text summarization, image generation and so on. GenAI leverages foundational Large Language Models (LLMs) that contains billions of parameters and are trained on vast amounts of data. These LLMs are able to learn intricate patterns and distributions to produce distinct outputs that resembles their training data, and generate human like text. Most of these LLMs, including OpenAI's GPT series, Meta's Llama series and Anthropic's Claude series~\cite{gpt,llama,claude} are build upon the Transformer~\cite{NIPS2017_3f5ee243} neural network, which is an improvement over the earlier RNN architectures. 

\textit{Agentic AI}~\cite{google-agentic} is a recent evolution that further advances the benefits of LLMs. An \textit{AI Agent}~\cite{aiagent1,aiagent2} is a software system that couples the decision making and generative capabilities of LLMs with \textit{actions} that can be performed based on the the recommendations of the LLMs. This gives agents autonomy to complete user-defined tasks. So, rather than just, say, generate code by prompting an LLM, an Agent using the LLM can compile and run the generated code and save the result in some location, i.e. LLMs form the 
``brain'' that thinks and instructs while the agent forms the ``body'' that can act upon this. These Agents are capable of reasoning, maintaining memory, adapting to environments and making dynamic decisions. Unlike traditional chat bots which follows a predefined deterministic script, these Agents can perform multi-step, complex and dynamic decisions. Recent agents even incorporate multi-modal capabilities, allowing them to perceive, process and produce various types of data including text, voice and video~\cite{aiagent1,aiagent2,aiagent3,aiagent4}. In this rapidly developing area, the capability of AI agents have further exploded through sophisticated \textit{prompting} strategies~\cite{prompting1,prompting2} and their capacity of using external \textit{tools}~\cite{tool-calling1,tool-calling2} as part of their actions. Tools allow Agents to interact with external systems including APIs, databases, search engines, etc., and such tools are exposed to the LLMs for them to be able to invoke  with specific parameters. 

The \textit{Model Context Protocol (MCP)}~\cite{mcp}, recently introduced by Anthrophic and being widely adopted, is an open protocol to standardize the description of tools/data sources and their parametrized invocation by LLMs. Prompting strategies help craft effective instructions and guidelines to the LLMs, which enhances the response outputs generated by them to effectively solve the task at hand. Techniques such as \textit{Chain-of-Thought (CoT)} prompting~\cite{cot}, where the model is encouraged to ``think aloud'' and break down problems into sequential steps, significantly enhances an agent's reasoning and problem-solving abilities. AI Agents are increasingly being used in real world applications, ranging from health care diagnoses~\cite{agentic-health} and software development~\cite{agentic-code}, to analyzing microscopy experiments~\cite{agentic-science} and writing a research article~\cite{agentic-article}.

\textit{Agentic Workflows} involve multiple Agents which work collaboratively to accomplish a user-defined task. Since AI Agents are autonomous by nature, Agentic Workflow executions are not prescriptive, and their invocation flow is non-deterministic allowing them to adapt to environment to solve complex tasks. Agentic workflows follows a generalized \textit{pattern of execution}, often involving steps like reasoning, planning, executing tools or actions, reflecting on outcomes, feedback and self correction loops~\cite{agentif-wf1,agentif-wf2}. Some popular design patterns include \textit{ReAct}~\cite{react}, \textit{Reflection}~\cite{reflection}, \textit{Reflexion}~\cite{reflexion}, \textit{Plan and Execute}~\cite{plan_and_execute} and \textit{Magentic-One}~\cite{magentic}. The rapid advancements of AI Agents and LLM models poses distinct challenges of seamlessly integrating these together, and with external data sources and tools. While seemingly powerful and able to replace humans for completing complex tasks, Agentic systems often struggle when faced with numerous tools and data sources, goals that require multi-step tasks with correct ordering, and long-context management to track critical information from the conversation history to avoid hallucinations. This motivates the need for improved \textit{agentic patterns} that allow the Agents to intelligently sequence steps and retain relevant information across multiple steps, avoiding both information overload and forgetting crucial details. Further, while protocols like MCP and \textit{Agent to Agent (A2A)}~\cite{a2a_protocol_2025} attempt to make these interactions less brittle, there lacks an architectural design to deploy and invoke \textit{MCP servers on cloud infrastructure} for operational use. This limits the real-world deployment, scalability and accessibility of MCP, beyond just sample developer applications on a desktop.

We address three key challenges in the emergent area of Agentic Workflows.
\begin{enumerate}[leftmargin=*]   
    \item We define a novel agentic workflow pattern, \textit{\agentx}, which is composed of stage designer, planner, and executor agents. The workflow is automatically generated based on decomposing the \textit{user prompt} that describes a task.
    \item We propose two alternative approaches for deploying MCP servers on cloud services, encapsulated as Function as a Service (FaaS), one using independent servers per tool and the another fusing them together.
    \item We perform detailed experiments of using the \agentx pattern for three realistic applications, for multiple instances of user prompts, and compare it against existing state-of-the-art (SOTA) agentic patterns, React and Magentic-One. We see that \agentx is competitive or better than them in progressing to the correct results for the tasks that we evaluate. We also offer detailed empirical evaluations on the two MCP deployment approaches and report their latency and cost, and also compare them with a local deployments.
    \item Lastly, we offer unique insights on the challenges encountered when designing and deploying Agentic Workflows for practical applications, highlighting several gaps between the promise of Agentic AI and practical limitations that require further principled investigation.
\end{enumerate}

The rest of the paper is organized as follows: we offer background and summarize related work~(\S~\ref{sec:bg}), describe our proposed agentic design pattern (\S~\ref{sec:x}) and MCP architecture (\S~\ref{sec:mcp-arch}), present comparative empirical results of our \agentx pattern and FaaS MCP architecture for several real tasks~(\S~\ref{sec:results}), discuss insights, observations and limitations of agentic workflows in the wild~(\S~\ref{sec:discuss}), and finally we present our conclusions and future work~(\S~\ref{sec:fw}). This work builds upon our prior workshop paper~\cite{ipdpsw} that set out preliminary ideas for the \agentx pattern, but improves upon those patterns and also introduces the MCP tool deployment and calling interfaces.

\section{Background and Related Work\pgs{1.5}}\label{sec:bg}

\subsection{Agentic Patterns}
Agentic design patterns are generalized execution flows for agentic workflows that involve steps such as reasoning, planning, executing tools or actions, reflecting on outcomes, providing feedback, and incorporating self-correction loops. These patterns acts as blueprints for structuring the behavior and interactions of AI agents. \textit{Reflection}~\cite{reflection} involves the AI model generating an initial response to a prompt, evaluating its own output for quality and correctness, and then refining the content based on its self-generated feedback. The \textit{Plan and Execute pattern}~\cite{plan_and_execute} facilitates human-AI collaboration on complex, multi-step tasks, where users and AI agents jointly compose and execute plans within a shared document. \textit{Magentic-One}~\cite{magentic} (Fig.~\ref{fig:magentic}(b)) represents a high-performing generalist multi-agent system that employs a multi-agent architecture. A lead agent, ``Orchestrator'', plans, tracks progress, re-plans to recover from errors, and delegates tasks to specialized agents such as a WebSurfer (for browsing), a FileSurfer (for file handling), a Coder (to generate code), and a Terminal (for shell execution). These patterns offer principled, repeatable and generalizable approaches for building complex, autonomous and collaborative AI systems. However, as we report in our experiments, these still have limitations in being able to solve non-trivial real-world tasks and our \agentx pattern improves upon these.

\subsection{Reasoning and Action}
Prompting strategies like \textit{Chain-of-Thought (CoT)} prompting~\cite{cot} have been instrumental in enabling LLMs to ``think aloud'' and find the right approach to solve a problem by breaking it into component tasks, significantly enhancing their problem-solving abilities. However, traditional CoT suffers from hallucination and error propagation due to its lack of access to the external world. The \textit{ReAct} paradigm~\cite{react} (Fig.~\ref{fig:react}(a)) uses LLMs to generate both reasoning traces and task-specific actions to interface with and gather information from external sources. Integrating the tool results into the reasoning process creates a powerful feedback loop, mitigating hallucination and error propagation. But it's use of a single context history causes hallucinations and high costs for multiple step tasks. Our \agentx pattern attempts to overcome these limitations. Preliminary ideas for the \agentx pattern that uses an evaluator and a judge to reason about the outcomes were introduced earlier~\cite{ipdpsw}, and the current article streamlines those patterns and system prompts based on subsequent experiments, and importantly introduces the MCP tool deployment onto FaaS and invokes the MCP interfaces for actions performed by the agents.

\subsection{Agentic Protocols}
Anthropic's MCP~\cite{mcp} and Google's A2A~\cite{a2a_protocol_2025} are recent protocols that have emerged to enable interoperable Agentic AI systems to interact between agents, LLMs, tools and data sources. MCP provides the LLM of the Agent with relevant information about the tools and standardized APIs for their execution using a client-server model. An \textit{MCP Host} refers to the Agent application which contains an \textit{MCP Client} that connects to an \textit{MCP Server}. The Server exposes APIs for the LLM to execute \textit{Tools}, retrieve Resources that are structured describing the tools, and fetch \textit{Prompts} that form reusable, user-defined templates. This gives better context for LLMs to make use of these tools in a more deterministic manner. 

Tools, and by extension the MCP servers which wrap around them, from an execution perspective can fall into three broad categories: remote, local and local-remote execution. \textit{Local MCP servers} have self contained executions when they are invoked, they do not depend or call external services or APIs to return the output of the tool called. An example of this in our application workflow discussed later is the \texttt{code-executor} MCP server, which receives the code from the agent and executes the code in our applications, that of plotting, locally. \textit{Remote servers}, on other hand, completely depend on the output on an external service for the results of the tool call, for example the \texttt{yfinance} MCP server is dependent on the Yahoo Finance servers to return the stock data of various companies. The MCP server, in this case, acts as a wrapper around the external service API, which improves the discoverability and usability of these services by agentic frameworks. \textit{Local-Remote servers} have a split execution profile where the MCP server is not just a wrapper around an external service but also performs remote and local executions to return an output for the tool call. An example is the RAG server, which depends on an OpenAI API for generating the embedding model but subsequently performs querying of the vector using an local in-memory store.

While ideally all MCP servers required will be hosted by an external service provider, given the nascent stage of MCP servers means that local and local-remote hosting of the servers will be necessary for the foreseeable future and to support unique tooling and privacy requirements of specific enterprises. Hence, from a systems perspective, local and local-remote servers are interesting to benchmark to help provide insights in how MCP servers hosted on cloud FaaS platforms can help improve latency, scalability and security of the agentic workflows, in contrast to running the MCP servers on the local machine executing the agents themselves. This is one of the goals of this article.

The A2A protocol enables interoperability among AI agents across different frameworks, and allows composition of complex workflows. It uses an \textit{Agent Card} to describe the skills and capabilities, security methods and supported data formats for agent discovery. While MCP enables an individual agent to interact with its environment, A2A provides the horizontal communication for inter-agent collaboration. LangChain~\cite{langchain} is also a popular framework that streamlines the development of LLM-powered applications by integrating retrieval, prompting, and external tools into modular, composable workflows. We focus mainly on the MCP protocol for integrating our agents with external sources and evaluating them, and leave A2A as future work.

\subsection{Serverless Cloud Computing}
\textit{Functions as a Service (FaaS)} is a serverless deployment and execution paradigm for modular business logic on clouds. Platforms AWS Lambda~\footnote{\href{https://docs.aws.amazon.com/lambda/}{AWS Lambda Docs}}, Azure Functions~\footnote{\href{https://azure.microsoft.com/en-us/products/functions/}{Azure Functions}} and Apache OpenWhisk~\footnote{\href{https://openwhisk.apache.org/}{Apache OpenWhisk}} allow users to write stateless functions and deploy these functions along with their dependencies on the cloud, scale their execution elastically with the load using containers, and offer granular billing per function call. Users do not have to launch server instances or allocate resources, easing manageability, and are billed for the duration of function execution. There are also FaaS workflows such as AWS Step Functions~\footnote{\href{https://aws.amazon.com/step-functions/}{AWS Step Function docs, https://aws.amazon.com/step-functions/}} and the cross-platform XFaaS~\cite{xfaas} that can help compose a collection of FaaS functions into a data flow graph for orchestration. FaaS offers a viable option for hosting one or more MCP servers in an scalable operational setting, and we propose different MCP hosting architectures using FaaS and their trade-offs in this article.

\begin{figure}[t!]

\centering
  \subfloat[ReAct]{
    \includegraphics[width=0.4\columnwidth]{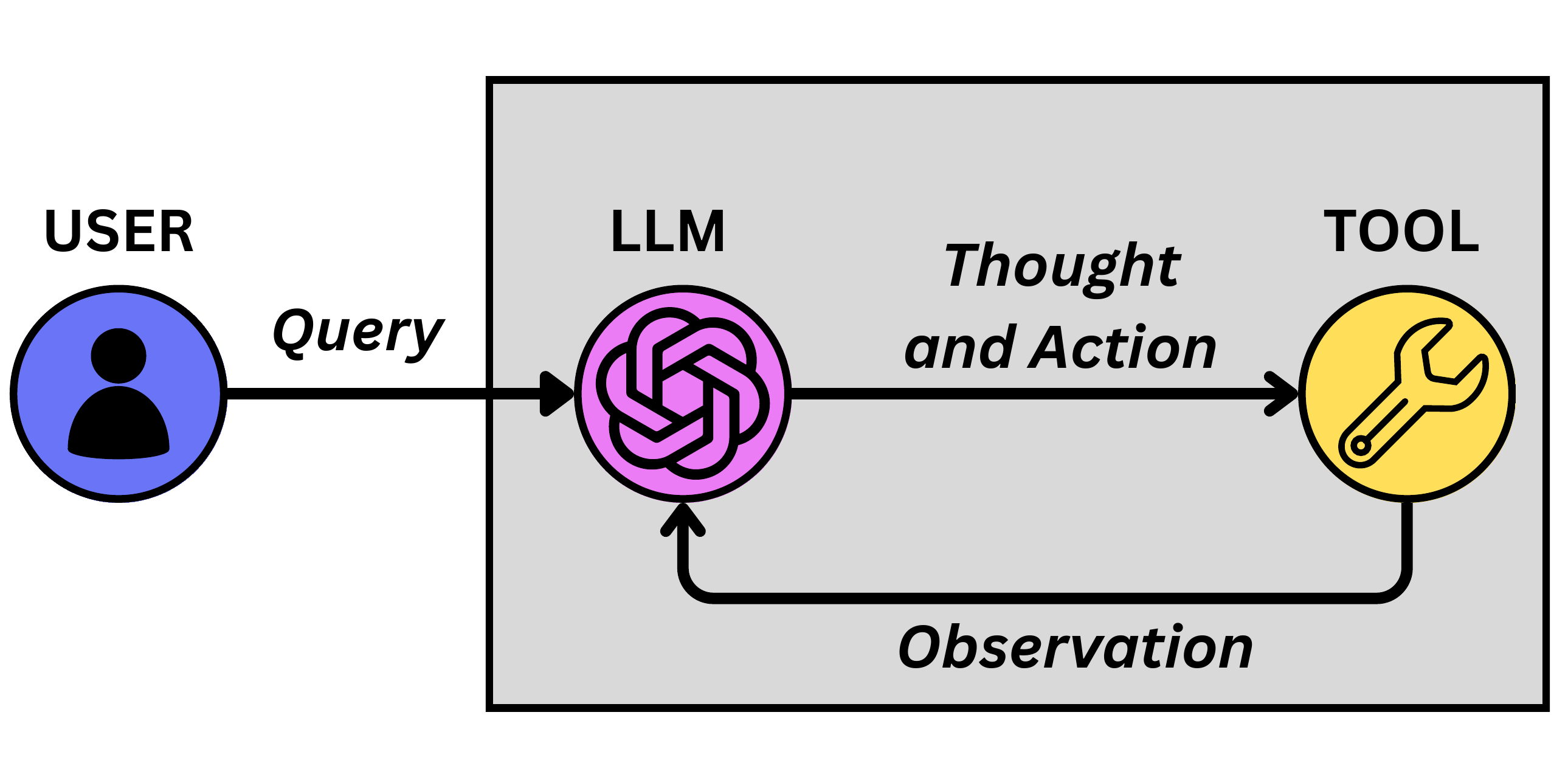}
    \label{fig:react}
  }
  \subfloat[Magentic One]{
    \includegraphics[width=0.5\columnwidth]{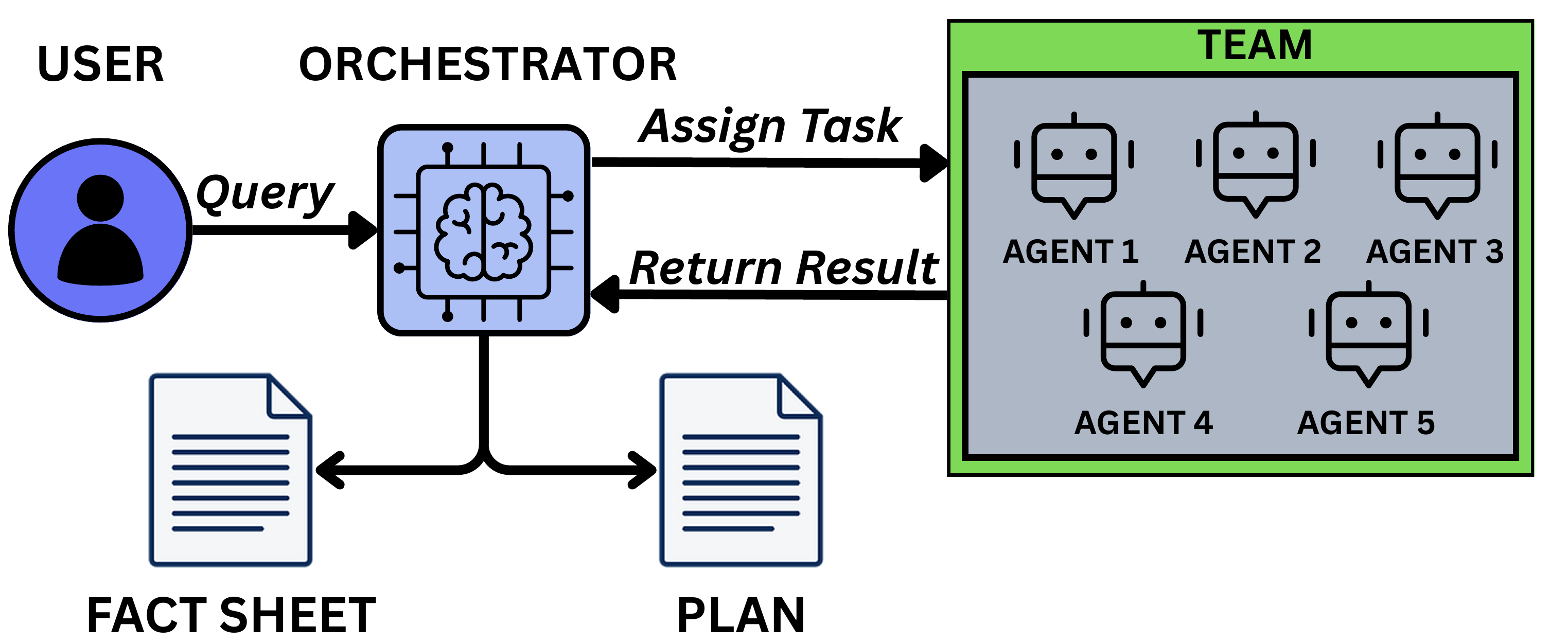}
    \label{fig:magentic}
  }\hfill
  \\
  \subfloat[AgentX]{
    \includegraphics[width=0.45\columnwidth]{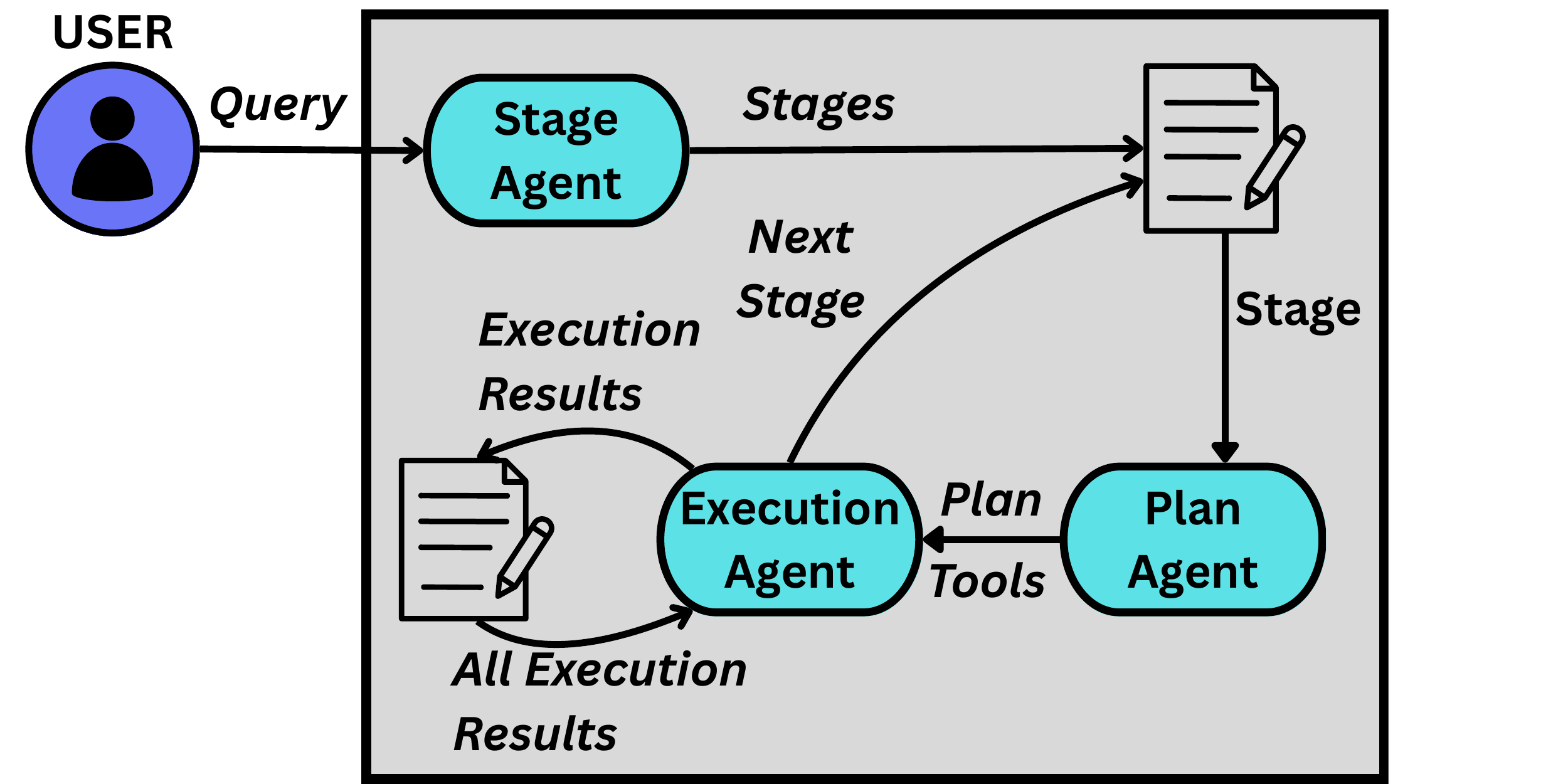}
    \label{fig:agentx}
  }

\caption{Agentic workflow patterns}

\end{figure}

\section{AgentX Workflow Design Patterns}\label{sec:x}
LLMs face a crucial challenge from their fixed context windows, which leads to a shortage of long-term memory capabilities and limits personalization during interactions with AI agents. This inherent limitation often disrupts the cognitive flow of extended interactions, requiring the re-establishment of context across sessions. \textit{Hallucination}~\cite{huang2025survey}, defined as the generation of LLM responses that are beyond their factual knowledge, poses a significant challenge to the reliability of chat bot systems and hinders their dependability in production environments. Mitigation strategies such as Retrieval-Augmented Generation (RAG)~\cite{rag} aim to ground LLM outputs in verifiable external sources. Despite advancements in prompting and tool use, Agentic Systems continue to struggle with a high degree of complexity arising from diverse external data and tools and intricate sequencing required for multi-step user tasks. To simplify development and ensure consistent, quality results, a more generalized execution flow is needed. This flow would effectively manage long-term memory and reliably sequence tools to handle extended interactions. 

\subsection{AgentX} 
We propose novel Agentic pattern, \agentx, composed of multiple agents that interact with each other (Fig.~\ref{fig:agentx}(c)): a \textit{Stage Generation Agent}, a \textit{Planner Agent}, and an \textit{Executor Agent}, designed to execute a generalized execution flow applicable to any application defined by a user prompt. Given a user task, the stage agent breaks down the task into a set of high level objectives called stages; the planner and the execution agents then processes each stage sequentially. The planner agent first makes a plan that contains detailed steps to achieve the objectives of the stage. The execution agent executes the stage while referring to the plan using a set of tools it has access to. The execution agent also generates execution results after each stage that is passed on as context for the next stage. The original user task is achieved once every stage has been executed to completion.

This design significantly simplifies the developer overheads while simultaneously offering robust quality to ensure progression and correct results, as shown empirically. By intelligently chaining the sequence of steps and selectively retaining relevant information across multiple steps, \agentx avoids both information overload and forgetting of crucial details, as we show from our higher success rate of completion, in \S~\ref{sec:results:latency:local-faas}. 

The agents in the \agentx pattern use of \textit{structured outputs}, which involves giving the agent an output schema that defines the structure of the output the agent should produce. The schema is provided as a Python object that includes attributes with a data type and description. e.g., the schema provided to the Stage Generator agent in \agentx has the \texttt{sub\_tasks} field with data type as a \textit{list of strings} and the description: \textit{``The list of sub tasks for the task''}. These schemas are automatically created by creating their corresponding pydantic class. Structured outputs help implement a flow of logic within the pattern by grounding the output of the LLM to a deterministic structure, which can later be parsed properly to extract relevant information for the execution flow. 

\subsection{Execution Sequence} When \agentx is given a task as a \textit{user prompt}, i.e., a high-level human description, its \textit{Stage Generation} Agent first breaks down the task into \textit{stages} or sub-tasks (Fig.~\ref{fig:agentx}(c)). These stages are then executed sequentially one at a time, where each stage includes the \textit{Planner} and the \textit{Execution} agents responsible for creating a plan for the stage and executing it. The user's task is completed when all the stages provided by the Stage Generation agent are completed successfully. This provides a generalized sequence of execution for any application, provided that the necessary tools are exposed to the LLM. Our design of the agents themselves involves the careful definition of their \textit{system prompts} and \textit{flow-pathways} that are exposed to each agent, to keep them focused on the task and improve the chances of completion. We describe these next.

\subsection{Stage Generation Agent} The \textit{Stage Generation Agent} breaks down the user prompt from a human-language description into a sequence of stages. We design this Agent with a \textit{system prompt} that specifies the guidelines to convert the given task into the \textit{least number of sub-tasks} required for an LLM agent with access to MCP tools to complete it such that similar or related sub-tasks are combined into a single sub-task when possible while also ensuring that the sub-task succeeds. The Stage agent also has the list of MCP tools descriptions from the MCP server or even the Doc String of a Python function as possible actions. This helps the Stage Agent create the stages considering the capabilities of the environment, and creating a sequence of stages that helps achieve the objective with least effort. The Stage Agent produces a list of stages, where each stage contains description of a high-level goal. Each stage is then passed to the Planner Agent for execution.

\subsection{Planner Agent} The \textit{Planner agent} creates a detailed plan for a given stage. We design the Planner with a system prompt explaining its role, which involves generating steps for a stage with their description, and also generating the exact tool and tool parameters to be used. The system prompt also instructors the agent to avoid redundancy in the steps created. The Planner is given the context of tools present in the system, the stages that have already completed, the current stage and the future stages, along with the original user prompt. By providing the stage context to the Planner we prevent it from making plans that solve already completed stages or future stages. This avoids duplication of effort across stages. 

The Planner creates a detailed plan consisting of a sequence of steps which guides the execution flow, and where the tools and their parameters are specified in the plan description. A key reason for this hierarchical planning is to provide the Execution agent with an execution plan that is not so large that it fails to complete it, while not too small to cause unnecessary token wastage. A key feature of the Planner is to expose \textit{only the necessary tools} relevant for the plan to the Execution agent; this helps the executor stick to the intended flow of execution and not be distracted by other available tools and hallucinate. This also reduces the overhead of tool descriptions, which becomes an issue when too many tools are exposed to the LLM and increases the input token cost. So, providing a plan with a list of steps that explicitly mentions the tool and the tool parameters to use increases the chances of success and efficient execution flow.

\subsection{Execution Agent} The \textit{Execution Agent} executes the plan made by the Planner agent. We design system prompts to first ensures that this agent \textit{understands} the plan and \textit{calls} the tools, and subsequently \textit{reflects} on the output of the tools. The system prompt for the first part of the agent execution simply states \textit{``Execute the following plan:...''} followed by the plan provided by the Planner. The execution agent loops over the tool calls and a reflection on its outputs as many times as required to successfully execute the plan. When tools are specified to be executed by the Agent, our \agentx framework invokes the tool, e.g., by trigger a local function call or by creating an MCP client to interact with the MCP server which wraps the tool. This mapping of tool name to implementation is maintained by our execution framework. 

During the reflection phase, the execution agent is passed two attributes in its output schema: the \textit{execution results} and a \textit{flag} to denote if the plan has successfully executed or not. In this phase, the execution agent is asked to summarize \textit{only the relevant information} from the stage to be passed to future stages. These results are later saved in a local variable which is used to provide context to the first phase of the execution agent. Tools that outputs a lot of content, like search results from a search engine, may contain superfluous content that may not be necessary for the stage but fills up the LLM's context window. By summarizing and providing only the necessary results from each stage to the execution agent of the next stage, we avoid this.

This memory consolidation to avoid bloated context by \agentx is only possible by breaking the user prompt into stages. This is also responsible for contributing towards reducing the overall cost of execution. Efficient context management also becomes vital when dealing with smaller LLM models deployed on edge devices.

\subsection{Distinctions of \agentx with SOTA Patterns}
Many agentic design patterns like \textit{ReAct}~\cite{react}, \textit{Reflection}~\cite{reflection}, \textit{Reflexion}~\cite{reflexion}, \textit{Plan and Execute}~\cite{plan_and_execute}, and \textit{Magentic-One}~\cite{magentic} have distinct strategies. \agentx offers a structured approach that uses hierarchical planning and context management.

The \textit{ReAct pattern} (Fig.~\ref{fig:react}) synergies reasoning and acting by prompting an LLM to generate interleaved ``Thought'' and ``Action'' steps~\cite{react}. This allows the agent to dynamically update its plan based on observations from external tools (e.g., API calls), effectively grounding its reasoning in real data and reducing hallucination. Its primary strength lies in interactive information gathering and multi-step problem-solving where the plan needs to be adjusted on the fly. However, ReAct's linear thought-action loop can struggle with long-horizon tasks because it lacks explicit mechanisms for long-term memory or context summarization which may result in higher costs.

Systems like \textit{Magentic-One}~\cite{magentic} (Fig.~\ref{fig:magentic}) employ a team of specialized agents coordinated by a central \textit{Orchestrator}. The Orchestrator creates a high-level plan and delegates sub-tasks to agents with specific tool-use capabilities. This design is highly versatile for open-ended tasks requiring diverse set of skills. The main limitations of Magentic one is its high resource consumption where it often requires dozens of LLM calls which leads to high latency and cost. Furthermore, while the Orchestrator plans and passes through a ``fact sheet'' as the downstream context, this can further be made robust as it fails to pass necessary context in some of our experiments.

Other patterns focus on improving the quality of the output through iterations. \textit{Self-Refine}~\cite{reflection} uses an LLM to generate, critique, and refine its own output in a loop until a stopping condition, boosting quality without any external data or finetuning however it is heavily reliant on the base model's ability to self-critique effectively. \textit{Reflexion}~\cite{reflexion} implements `'verbal reinforcement learning'' where the agent reflects on past failures observed over multiple runs to improve in subsequent trials. While powerful for enhancing accuracy and enabling learning from mistakes across multiple runs, it is computationally expensive due to their iterative nature and does not guarantee recovery due to its reliance on the base model's ability to identify and improve a failure.

\agentx makes use of concepts from some of these patterns, and adopts a novel and structured workflow (Fig.~\ref{fig:agentx}) with several key distinctions:
\begin{itemize}
    \item \textbf{Hierarchical Planning:} Unlike the single-level planning in Magentic-One or the interleaved planning in ReAct, \agentx employs a two-tier hierarchy where the \textit{Stage Generation Agent} first creates a high-level, coarse grained plan (stages), and then the \textit{Planner Agent} creates a detailed, fine grained plan for each stage. This decomposition simplifies complex problems and improves robustness.
    \item \textbf{Strict Agent Roles and Tool Filtering:} \agentx enforces specialized, functional roles (\textit{Stage Generation}, \textit{Planner}, \textit{Executor}) rather than tool-based specializations like Magentic-One or using a single agent for both reasoning and action like ReAct. The Planner agent filters and provides \textit{only the necessary tools} for a given stage to the Executor. This aims to minimize the Executor's cognitive load, reducing the risk of hallucinating incorrect tool usage, and lowers token costs by shrinking the prompt context which adds up quickly when dealing with numerous tools.
    \item \textbf{Active Context Optimization:} \agentx focuses explicitly on memory consolidation. After each stage, the \textit{Execution Agent} reflects on the tool outputs and summarizes, keeping only the relevant information to be passed as context to the next stage. This prevents the context window from becoming bloated with too many details which is a common issue in long-running tasks.
\end{itemize}

\section{MCP Deployment Architecture}
\label{sec:mcp-arch}
\begin{figure}[t!]
\centering
  \subfloat[Local MCP Servers]{
    \includegraphics[width=0.25\columnwidth]{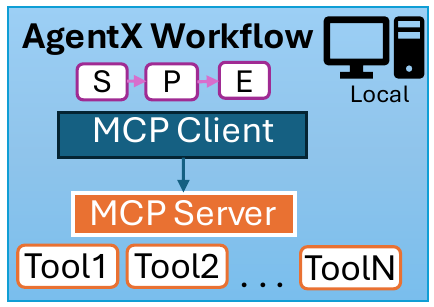}
    \label{fig:mcp_local}
  }
  \subfloat[Monolithic FaaS MCP]{
    \includegraphics[width=0.25\columnwidth]{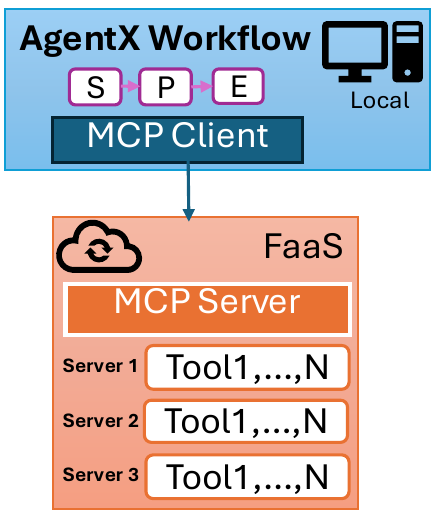}
    \label{fig:mcp_monolithic}
  }
  \subfloat[Distributed FaaS MCP]{
    \includegraphics[width=0.45\columnwidth]{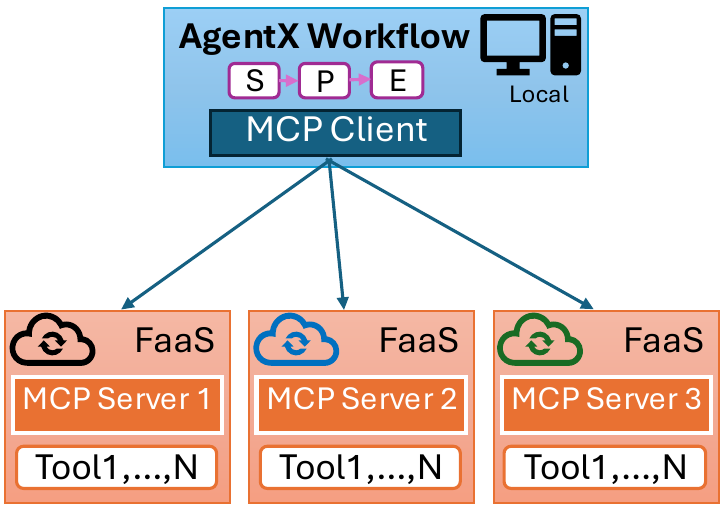}
    \label{fig:mcp_distributed}
  }
\caption{MCP Deployment Architectures}
\label{fig:mcp_all}
\end{figure}

MCP is designed as a stateful, long-running, and context-aware server architecture that exposes services accessible to agentic workflows. Agentic frameworks usually execute on the local machine of the user and are configured to access multiple external LLM endpoints (e.g., OpenAI, Claude) as well as local MCP servers (Fig.~\ref{fig:mcp_local}) that are required to complete various user tasks. However, running the MCP servers locally puts the responsibility for maintaining tooling and their dependencies on the agentic workflow developers rather than the managers of the tools and services exposed by the MCP server. While MCP servers provide greater structure for tool invocations by agents, they still require installation and framework-specific configuration (Fig.~\ref{fig:mcp_local}). This introduces developer and operator overheads during the development and execution of agentic applications. 

In this paper, we propose multiple alternative deployment models for MCP servers onto a cloud-hosted FaaS platform. We evaluate these using AWS Lambda, though the design itself is FaaS service-provider agnostic. A FaaS-hosted MCP server offers improved security, scalability and ease of access for users of the server within an enterprise, relative to a local MCP server execution, while retaining control over the server and its data regulations by the enterprise. We explore two distinct FaaS MCP deployment approaches: a \textit{monolithic} function, encapsulating multiple MCP servers (Fig.~\ref{fig:mcp_monolithic}) and a \textit{distributed} function design that hosts each MCP server as a separate function (Fig.~\ref{fig:mcp_distributed}).

The monolithic approach (Fig.~\ref{fig:mcp_monolithic}) requires just a single FaaS deployment and simplifies management, but will require changes to the function if additional MCP servers are added, incur higher costs per function execution due to larger memory footprint, introduce potential latencies, and risk overloading the agent using the service with unnecessary MCP servers. Conversely, the distributed architecture (Fig.~\ref{fig:mcp_distributed}) provides more granular deployment of MCP servers that can be reused, with only a subset provisioned to an agent, and lowers the memory footprint of the function but adds additional deployment complexity of multiple functions. This represents a trade-off that needs a systematic exploration. In this article, we evaluate the distributed MCP deployment model and leave the experimental comparison with a monolithic approach as future work.

\subsection{MCP Servers Evaluated}
Table~\ref{tab:tools} lists the MCP servers that the agentic frameworks can utilize to complete the three applications we evaluate in our experiments. Each MCP server exposes one or more tools, resources and/or prompts. These are Python or Node.js servers sourced from the official MCP GitHub repository~\cite{mcp-github}. We have implemented additional MCP servers when they were found essential for the application and where an open-source implementation was not already available. While some servers execute tools exclusively locally (on the local machine or the FaaS container hosting the server), others make additional remote API calls.

The \textit{Tools} column of Table~\ref{tab:tools} indicates the distinct tools exposed by each server that agents can invoke. The \textit{Origin} column mentions if the server was officially released by the original tool provider, is a community-supported open source implementation accessing external (official) APIs, or were custom developed by us. For instance, the \textit{Code Execution} server provides a Python environment for agents to execute scripts, the \textit{yFinance} server scrapes the Yahoo Finance website for stock data, while the \textit{Serper} server enables web searches using the Google Serper API\footnote{\href{https://serper.dev/}{Serper API}}. The \textit{arXiv} server queries and downloads research articles from arXiv, \textit{Fetch} enables agents to scrape webpages for content given their URL, and the \textit{File System} server enables seamless read/write operations with the local file system. However, this server is not employed in the FaaS experiments as Lambda functions lack persistent local storage. Instead, we develop a custom \textit{S3} MCP server to read and write objects to S3, serving as an analogue to the File System server in FaaS experiments. We also implement the \textit{RAG} server as it is crucial for information retrieval from unstructured text data such as papers.

\subsection{Distributed MCP Implementation on AWS Lambda} 
FaaS-based MCP servers in this article use the distributed deployment approach (Fig.~\ref{fig:mcp_distributed}(b)). These are currently implemented as AWS Lambda functions but can be replaced by equivalent Azure Functions or Google Functions, or hybrid cloud deployments as future work. All servers used in our application workflows have been containerized. Such a dockerised deployments overcomes the storage limits of in-line AWS Lambda functions and provides support for function requiring up to 10GB, which is particularly important when deploying resource-intensive servers. E.g., the \textit{code-executor} and \textit{yfinance} servers require either a large number of supporting packages or substantial file dependencies.

The servers are deployed on AWS Lambda with the least memory allocation required to execute them, as shown in Table~\ref{tab:tools}. Since FaaS functions are billed by memory used and duration of execution per call, provisioning less memory for the function lowers the invocation costs. All Lambda functions are provided with a standard ephemeral storage of 512MB. The Lambda Function URL maps incoming HTTP requests to the event object and invokes the Lambda function handler, which internally leverages the \textit{mcp-lambda-handler} module from the \textit{awslabs} GitHub repository\cite{mcp-lambda-handler}, to process events such as listing and invocation of tools.

The MCP servers are invoked directly via AWS Lambda Function URLs which have dedicated HTTPS endpoints that provide simple, low-latency HTTP interface without API Gateway features like advanced routing or throttling. MCP Clients call the HTTPS Function URL, which Lambda maps to an event; the Lambda MPC handler then translates that HTTP event into a JSON-RPC request as per the MCP standard, invokes the registered tool or resource with the MCP server, and serializes the JSON-RPC result back to the client as the HTTPS response.

\textit{Statefulness} (session management) is necessary to track any local files stored in the \texttt{/tmp} folder for Lambda function instances across multiple calls to the MCP servers but as part of the same agent interaction session. This can however be extended to any application requiring statefulness across MCP invocations. We implement statefulness in FaaS using a \textit{session\_id} that is returned in the server's HTTP response and persisted in AWS DynamoDB for use in subsequent requests. An \texttt{INITIALIZE} request is sent each MCP server at the start of each application interaction session to create a \textit{session\_id} for that specific application instance. The unique \textit{session\_id} is created by each MCP server for that application instance, ensuring isolation between concurrent agent operations for different applications. All the agents for an application instance subsequently use these generated \textit{session\_id} for invocations of the MCP servers. Finally, as the application completes, all servers receive a \texttt{DELETE} request that removes the sessions from the DynamoDB table.

For the FaaS deployment, the MCP tool descriptions are minimally modified to ensure they can successfully run in Lambda. E.g., we instruct them to avoid using local container storage in the FaaS deployment and instead use S3, done by appending to the end of each user query the additional prompt: \textit{``...you can read/write from s3 from this location: `\texttt{s3://dummy-bucket/agent/}' ''}, or if it involves only writing we append the additional prompt: \textit{``...write it to s3 location: `\texttt{s3://dummy-bucket/agent/}' ''}. Further, the Document Retriever tool in RAG is modified to ensure that agents provide the S3 URI to the tool to read the ArXiv PDF file rather than a local path used in the non-FaaS deployment, given as: \{~\textit{\underline{Description:} Retrieves relevant text snippets from a PDF in S3 based on a query. \underline{Input:} \texttt{s3\_uri (str)}: The S3 URI to the PDF file (e.g., \texttt{s3://my-bucket/report.pdf}). \texttt{query (str)}: The query to search in the PDF file. \underline{Output:} \texttt{str}: Snippets of text from the PDF relevant to the query, with metrics.}~\}

\begin{table}[t!]
\centering
\small
\renewcommand{\arraystretch}{1.0}
\caption{MCP Server Descriptions}
\label{tab:tools}
\begin{tabular*}{\textwidth}{@{\extracolsep{\fill}}l c l c c c@{}}
\toprule
\textbf{MCP Server} & \textbf{Tools} & \textbf{Origin} & \textbf{Execution} & \textbf{Mem. (MB)} & \textbf{Storage (MB)} \\
\midrule
Code Execution & 4 & Custom & Local & 512 & 512 \\
RAG & 1 & Custom & Remote & 512 & 512 \\
YFinance & 17 & Community & Remote & 128 & 512 \\
Serper & 13 & Community & Remote & 512 & 512\\
Arxiv & 8 & Community & Remote & 256 & 512 \\
Fetch & 9 & Official & Remote & 256 & 512 \\
File System & 10 & Official & Local & N/A & N/A \\
S3 & 3 & Custom & Local & 128 & 512 \\
\bottomrule
\end{tabular*}
\end{table}

\section{Experiments and Results\pgs{4.75}} \label{sec:results}

\subsection{Setup}
We compare the performance of \agentx with two other SOTA agentic patterns as baselines: \textit{ReAct}~\cite{react} and \textit{Microsoft's Magentic-One}~\cite{magentic} (Fig.~\ref{fig:magentic}). The ReAct pattern is implemented using \textit{LangGraph}~\cite{langgraph}. We use the \textit{create\_react\_agent} function and obtain the execution trace from the observability tool, \textit{LangSmith}. The ReAct agent we use is a divergence from the original ReAct pattern, and consists only of the action and observation components, omitting the thought component. While LangGraph supports the classic ReAct pattern, that implementation is limited to only using tools with a single parameter. Most of the MCP servers used in our experiments involve tools with multiple input parameters. 
The Magentic-One pattern is implemented using \textit{Microsoft's AutoGen framework}~\cite{microsoft-autogen,wu2024autogen}, and the execution trace is obtained using the observability tool, \textit{AgentOps}~\cite{agentops}. The original implementation of Magnetic One features a team based architecture composed of four specialized agents: a file system agent, a web surfer agent, a coder agent, and a computer terminal agent. By default, these agents operate with a predefined set of tools rather than utilizing MCP servers. In our experimental setup, we adapt this architecture by replacing the default agents with a new set of agents, creating a distinct agent for each MCP server. A key modification is the addition of a  descriptions for each new agent, outlining its specializations, capabilities, and intended purpose. This is necessary because the orchestrator in Magnetic One relies on these descriptions to effectively delegate tasks and create plans. For example, the agent given access to the Arxiv MCP server is designated as the Arxiv Agent and is assigned the following description: \textit{``Agent for interacting with the arXiv API to retrieve article URLs, download research papers as PDFs, load articles into context, get article metadata, and perform search queries on arXiv.org.''}.

We implement our \agentx pattern using a Python framework consisting of modules for the different agent types and an orchestrator between the agents to implement the interaction pattern. It does not depend on external frameworks. The execution trace is returned by each module. For all our experiments, we utilized OpenAI's gpt-4o-mini large language model, accessed via their public cloud API. We used the specific model version gpt-4o-mini-2024-07-18 with a context window of 128,000 tokens. However, its parameter count has not been publicly disclosed. All API calls were made to the Chat Completions endpoint using the official openai Python library (v1.35.0). The model was prompted with a conversation history, provided as a list of objects to the messages parameter, and a set of available functions, defined as JSON schema objects within the tools parameter. To ensure consistency and reproducible outputs, the temperature and top\_p were set to their default value of 1.0.

All patterns and their agents execute on a Dell Precision 3440 with Intel Core i5-10500 processor (6 cores, 12 threads) and 16 GB RAM, running Ubuntu 20.04.6 LTS (64-bit). For the local MCP variant (Fig.~\ref{fig:mcp_local}), the MCP servers also run on the same workstation. For the experiments with FaaS MCP servers, we deploy the MCP servers on AWS Lambda using the distributed approach (Fig.~\ref{fig:mcp_distributed}), as described above. Since lambda functions do not support run time installations of python packages, all the dependencies for each MCP server are preinstalled. For example, for the \textit{Code Execution} MCP server, all the packages required to execute a standard plotting script including \textit{matplotlib} and \textit{pandas} are preinstalled.  

Each MCP server acts as a wrapper around its tools, serving as the endpoint with which agents on the local workstation can invoke tools. In our experiments, we compare the local MCP server variant with the distributed FaaS variant, for the same application workloads, and analyze them across different dimensions including end-to-end response time, input and output tokens. 

\subsection{MCP Server Configuration}
Tool descriptions play a major role in deciding the flow of execution for an agentic workflow. The tool is advertised by the MCP server by including its intended use, advantages, disadvantages, and expected input and the output it provides. Based on this, the LLM of the agent would choose to use the tool or not, and generate the parameters to the tool. 

The default tool descriptions provided by the MCP servers sometimes contain inadequate information that causes the LLM to not use the tool for its intended purpose. E.g., the default description for the Fetch tool is: \textit{``Fetches a URL from the internet and optionally extracts its contents as markdown.''}, which explains the extraction of content from the URL, but no pattern ends up using this tool for the web search application.

To encourage the use of the Fetch tool in the application executions, we append the following additional hint to the tool description: \textit{``Use this tool after using the Google Search tool, when you need more detailed information from a specific web page.''}. This ensures that the agentic patterns use the Fetch tool and the intended flow of execution is followed, leading to more meaningful outcomes. A similar behavior was found for the Load article to context tool in the Arxiv MCP server where the default tool description: \textit{``Load the article hosted on arXiv.org into context ...''} causes the patterns to load the entire article as text into their context, which can exceed the maximum input token limit for the GPT 4o mini. So, we append this to their tool description: \textit{``This tool should never be used to load research papers since they are too long.''} to ensure that the agentic patterns do not use this but instead the Document retriever tool for querying information from a document. These hints have been added only for the Local set of experiments and not FaaS MCP servers. 

Also, the FaaS MCP servers only host a subset of the tools that are present in the MCP servers hosted locally. This is because the MCP servers cannot directly be deployed on FaaS and need some changes. E.g., tools that require multi-threading do not function correctly in the FaaS environment. Therefore, only the tools that are relevant for an application's intended execution flow are preserved in the MCP servers.

\subsection{Application Workflows}
\label{sec:apps}

\begin{figure}[t!]
\centering
  \subfloat[Web Exploration]{
    \includegraphics[width=0.6\columnwidth]{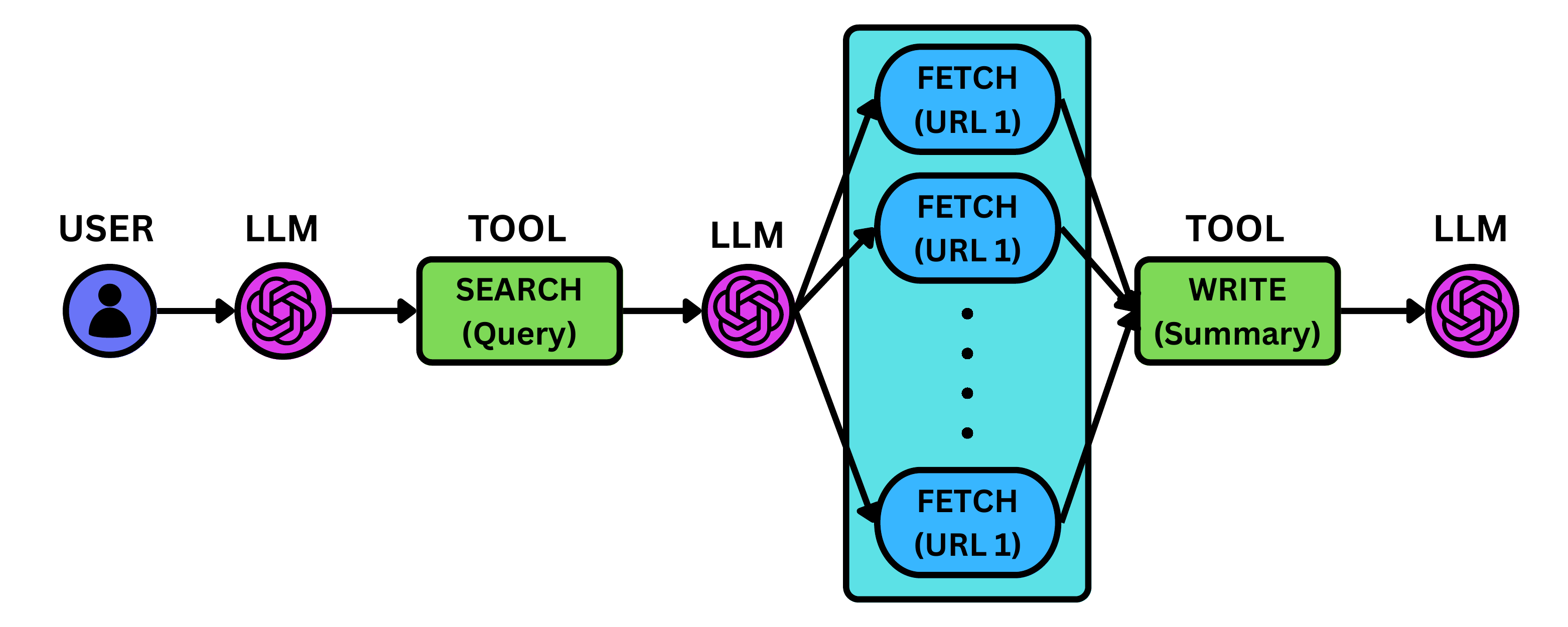}
    \label{fig:WebSearch}
  }
  
  \subfloat[Stock Correlation]{
    \includegraphics[width=0.4\columnwidth]{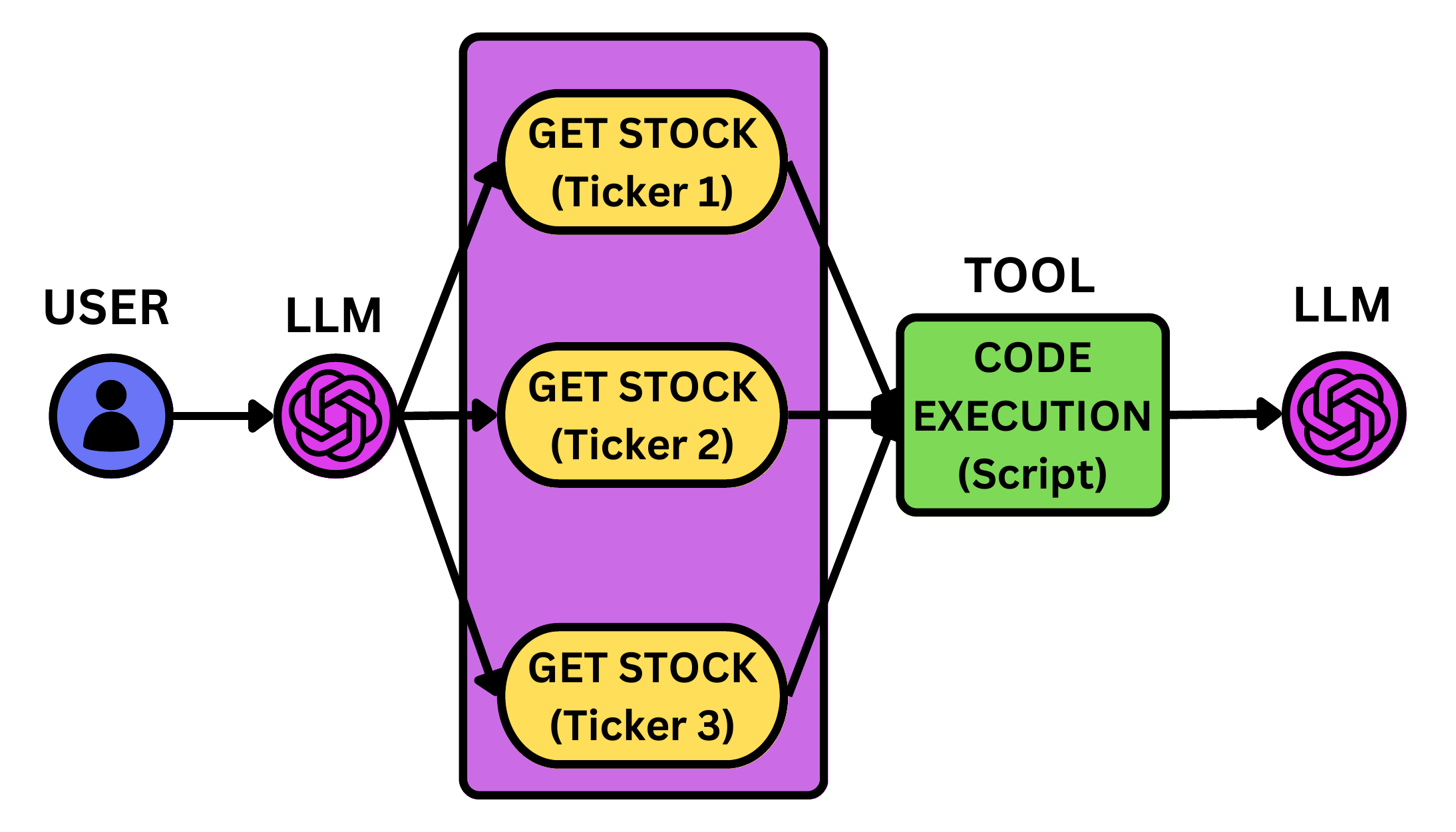}
    \label{fig:stock}
  }\hfil
  \subfloat[Research Report Generation]{
    \includegraphics[width=0.4\columnwidth]{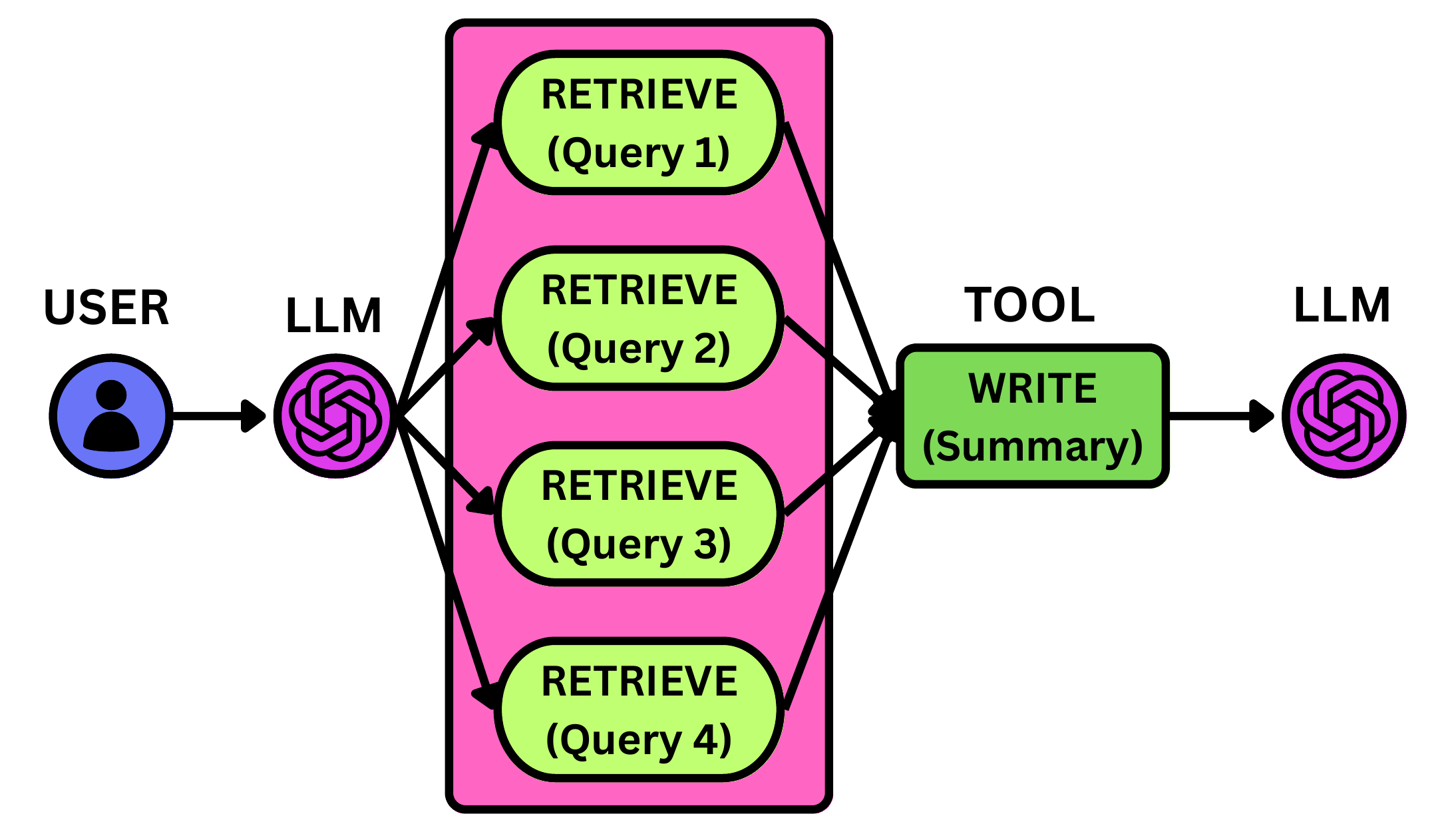}
    \label{fig:research}
  }
\caption{Expected workflow execution behavior of the three applications.}
\label{fig:apps-all}
\end{figure}

We specify three user application templates and perform experiments using all three patterns for these applications. Each application accesses one or more MCP servers, each exposing multiple related tools (Fig.~\ref{fig:apps-all}). The applications are templatized such that they can be used for different instances, and we use three \textit{instances} for each application using different variables for the template. For example, the research report application is templatized to generate reports on various \textit{topics}, which serve as placeholders, and we use phrases relevant to quantum computing and materials research for these topics. This enables robust comparison across patterns by eliminating chance-effects where a pattern performs better or worse for one specific topic rather than others. It also provides a more comprehensive understanding of how these patterns perform for the application as a whole in a generic manner. We also perform $\approx 5$ \textit{runs} for each instance of an application.

\subsubsection{Web Exploration} 
A common use-case of LLM powered agents is to navigate the web and extract necessary information. This application involves the Agent searching the web using tools for retrieving relevant URLs for a topic and summarizing the text content from those URLs. The web exploration application template has the simple user prompt: \textit{``Search for [query] and summarize the results in a text file''}, with \textit{[query]} being replaced with the following instances: \textit{`Recent advancements in quantum computing hardware development'}, \textit{`Edge devices and their real-world use cases in 2025'} and \textit{`Latest trends in biodegradable materials for sustainable packaging'}. The expected workflow of execution flow for this application is shown in Fig.~\ref{fig:WebSearch}. However, since the workflow is not imperatively define but automatically generated by the agentic pattern based on the user prompt, the actually workflow that is executed can depend on the specific agentic pattern or even the application or its instance. These are discuss in the results.

\subsubsection{Stock Correlation} This application involves the LLM gathering the price history of multiple stocks, and compare and correlate them generating a visualization/plot. It has the user prompt template: \textit{``Generate a plot for the historic stock prices of [A], [B], and [C] and save it as $[A][B][C].png$.''} with the stock triples \textit{`[A],[B],[C]'} having the instances \textit{`Apple, Alphabet (Google), Microsoft'}, \textit{`Netflix, Disney Amazon'} and \textit{`Coca-Cola, PepsiCo, Mondelez'}. The Agent is provided with MCP tools to gather these historic trends while also providing it with a Python code execution environment for executing Python scripts. While the workflow should involve two basic steps, gathering stock information and plotting them as shown in Fig.~\ref{fig:stock}, it provides a case of how hallucinations, context switching and lack of proper planning can effect the execution flow and cause the agents to take circuitous pathways.

\subsubsection{Research Paper Summarization} This application searches for and summarizes a research paper. The agent is provided MCP tools to search and download research papers in PDF format from arXiv. Since loading all documents into a single LLM query is not feasible for large research articles, we also provide the agent with a custom RAG MCP Tool that we have developed~\cite{rag}. This server splits the document into overlapping chunks and computes their embeddings using an external embeddings API, OpenAI’s \textit{text-embedding-3-large}, and stored in an in‑memory vector store initialized within the MCP server. At query time, the user query is embedded using the same API and compared with the stored embeddings using similarity search; snippets whose scores exceed a predefined threshold are retrieved for downstream generation.

The research report generation application template uses the user prompt: \textit{``Generate a report on the Core Contributions, Methodology, Experimental Results, and Limitations for the paper titled [title] and save it as a text file.''}, with \textit{[title]} being one of these instances: \textit{`Why Do Multi-Agent LLM Systems Fail?'}, \textit{`Flow: Modularized Agentic Workflow Automation'}, and \textit{`Magentic-One: A Generalist Multi-Agent System for Solving Complex Tasks.'}. In this workflow (Fig.~\ref{fig:research}), the agent is expected to issue multiple queries about the given paper title to produce a summary and is granted tools both to download and load the paper’s content and to perform RAG over the ingested text. This provides the agent with multiple pathways to complete the application and generate the desired results.

\subsection{Results and Analysis} 
We provide below a detailed analysis of the empirical comparison of these three applications using the three agentic patterns, and the local and FaaS MCP hosting, for diverse metrics, along with reasoning for the results obtained. Later, the Anomalies  sub-section in the discussion delves further into unique observations in these results. In the plots, \textbf{``R''} refers to \textit{ReAct}, \textbf{``A''} refers to \textit{\agentx} and \textbf{``M''} indicates \textit{Magentic One}. For each application, Web Search, Research Report and Stock Correlation, we report the instances identified by their placeholder variable, e.g., ``Quantum'', ``Edge''' and ``Materials'' indicating the topics used for the Web Search application.

\subsubsection{Analysis of Accuracy of Generated Results for Applications}\label{sec:results:accuracy}
The average accuracy of different agentic patterns are calculated for the the three applications across their three instances. The accuracy score is calculated using a set of attributes tailored to the application. These scores were then aggregated to generate a final score. The final output generated for each execution of an instance by a pattern is passed on to an OpenAI 4o-mini model with specific system prompts based on the attributes discussed below, and scored out of 100 for each attribute. Finally, we weight the score for an attribute based on its importance. For example, Accuracy and Relevance are most critical for content quality compared to Depth and Breadth of the content covered in the summary. Similarly, Data Accuracy and Query Adherence  are more important compared to the quantity of data and plot quality. The average of the weighted sum of the scores gives the final score out of 100. 

Web Search and Research Report generate a summary, and therefore we use the same set of attributes to evaluate them: Accuracy, Relevance, Depth, and Breadth. Accuracy measures the factual correctness of the content, ensuring that claims and details are verifiable, free from hallucinations and are logically coherent; this attribute was assigned a weight of 50 out of 100 due to its central role in determining trustworthiness. Relevance assesses how closely the content aligns with the user’s query and avoids irrelevant information, it is assigned with a weight of 30. Depth captures the extent to which the summary moves beyond surface-level statements to provide explanations, context, and detail; it is assigned a weight of 10. Breadth reflects the coverage of all major sub-topics and perspectives relevant to the prompt, and is also weighted at 10. Depth and Breadth, while important, are secondary to correctness and focus, and hence carry lower weights.

For Stock Correlation, which requires the generation of executable code that produces a plot, have a different set of attributes defined. Data Accuracy evaluates whether genuine historical stock data is used, and is the most heavily weighted attribute at 50 similar to accuracy above. Query Adherence measures how strictly the generated code follows the explicit requirements of the user’s prompt, such as including the requested companies, using the requested filenames and saving the plot, and carries a weight of 30. Plot Quality evaluates the clarity, readability, and professional presentation of the plot, with a weight of 10. Finally, Data Quantity considers whether the volume of data used is sufficient to create a meaningful plot, also with a weight of 10. Here again, correctness and adherence to the task dominate the evaluation, while presentation and supporting factors play a smaller role.

\begin{figure}[t]
\centering
\includegraphics[width=\columnwidth]{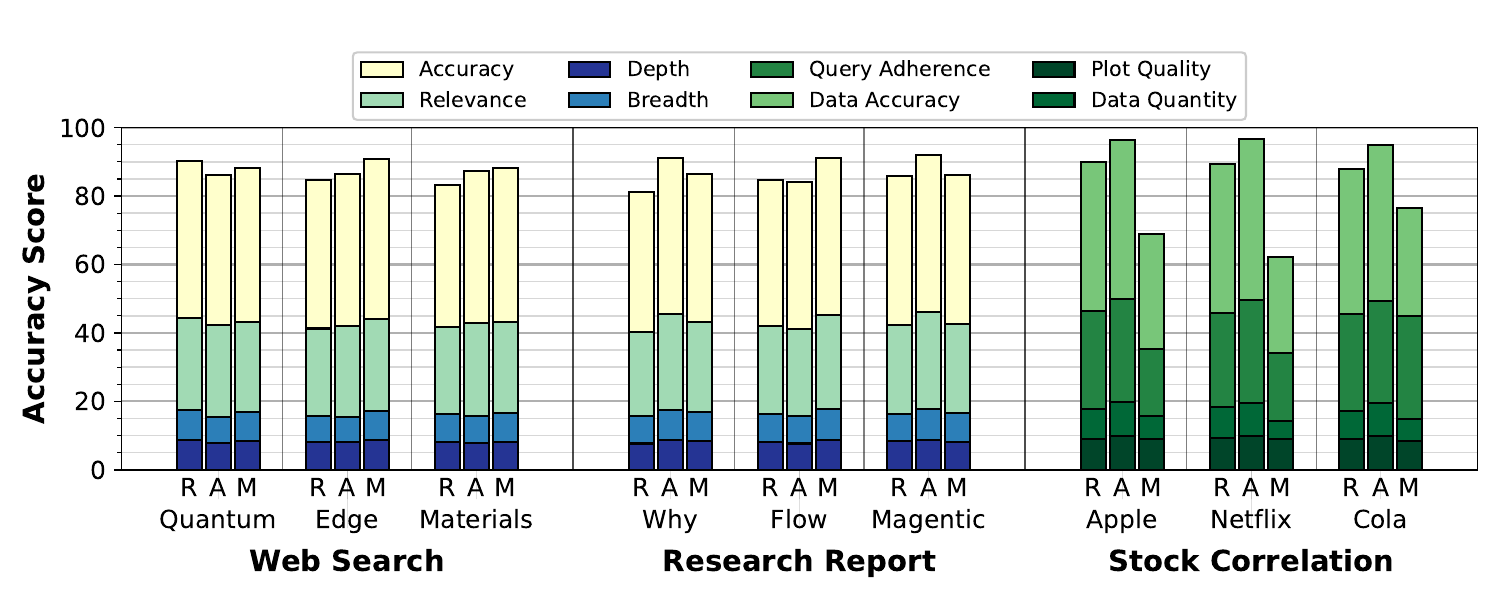}
\caption{Average accuracy score for all the runs.}
\label{fig:accuracy}
\end{figure}

Fig.~\ref{fig:accuracy} shows the accuracy score on 100 and contribution by each attributed for the patterns and application instances. We see that the accuracy of ReAct, AgentX and Magentic One for the web search and the research report application are similar, e.g., React, AgentX and Magentic One have an average score of 86.1, 86.7 and 89.1 for the Web Search, and  83.9, 89 and 87.9 for the Research Report application. This is because the patterns follow a similar set of steps to reach generate final output, causing the final outputs to be comparable to each other. For the Stock Correlation application, ReAct and AgentX have consistently higher scores than Magentic One, by 28.9\% and 38.8\%, respectively. Magentic One has significantly lower performance on 'Data Accuracy' and 'Query Adherence' attributes, where it has an average score of 64.3 and 78.7, and which have higher weights. This is because Magentic One truncates or even \textit{fabricates} the data when generating the plot code, which causes the scores to drop.

\subsubsection{End-to-End Application Latency Analysis}\label{sec:results:latency}

\begin{figure}[t]
\centering
\includegraphics[width=\columnwidth]{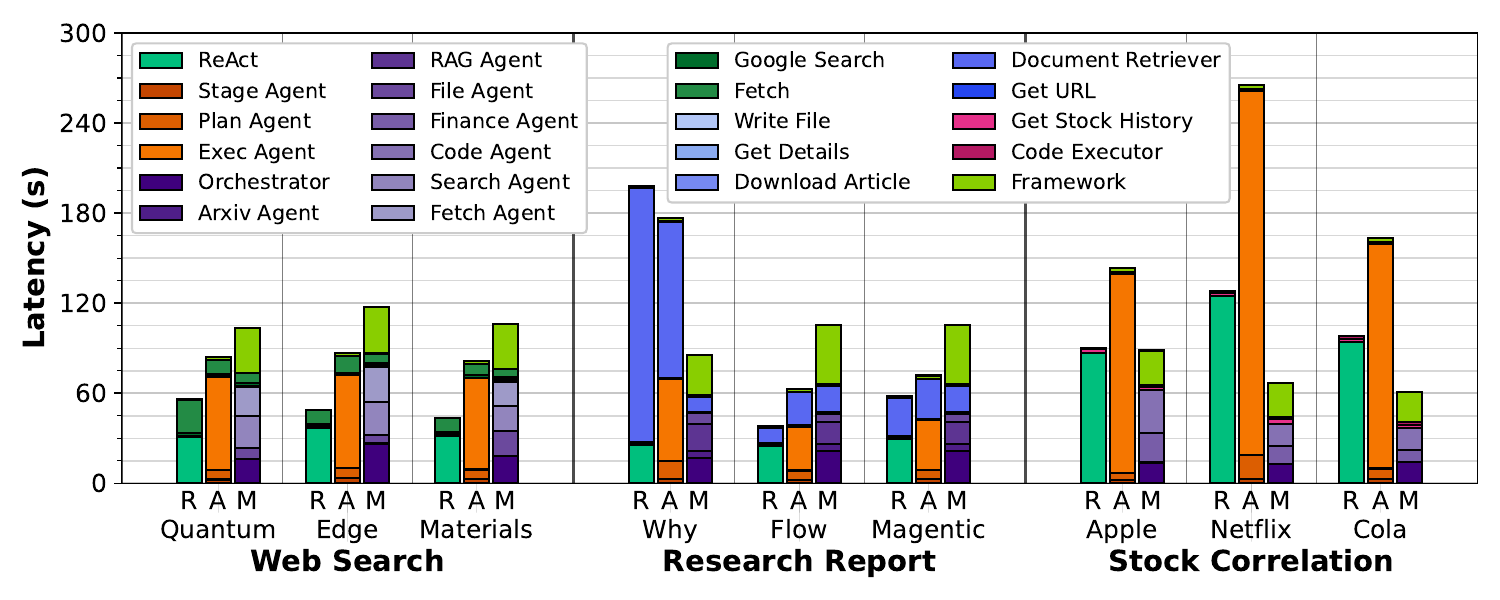}
\caption{Average latency to complete the application across 5 runs for local executions.}

\label{fig:Latency-Local}

\end{figure}

\paragraph{Local MCP}
A comparison of end-to-end application latency for local MCP execution across the three agentic frameworks show variable performance, as illustrated in Fig.~\ref{fig:Latency-Local}. The figure describes the latency contribution made by LLM inferences and tool invocations along with the framework latency through out the application's execution. Each of the colored stack belongs to either an agent belonging to one of the agentic patterns or a tool belonging to one of the application's intended execution flow or the framework latency. The framework latency is the remaining time obtained after removing agent and tool latencies from the total latency which involves many variables including network latencies and profiling latency. Agents belonging to the same agentic patterns are of different shades of the same primary color, e.g. Orchestrator, Arxiv Agent and File Agent are of different shades of purple. Similarly, Tools belonging to the same application are of different shades of the same primary color, e.g. Download Article and Document Retriever are of the different shades of blue. 

\begin{figure}[t]
\centering
\includegraphics[width=\columnwidth]{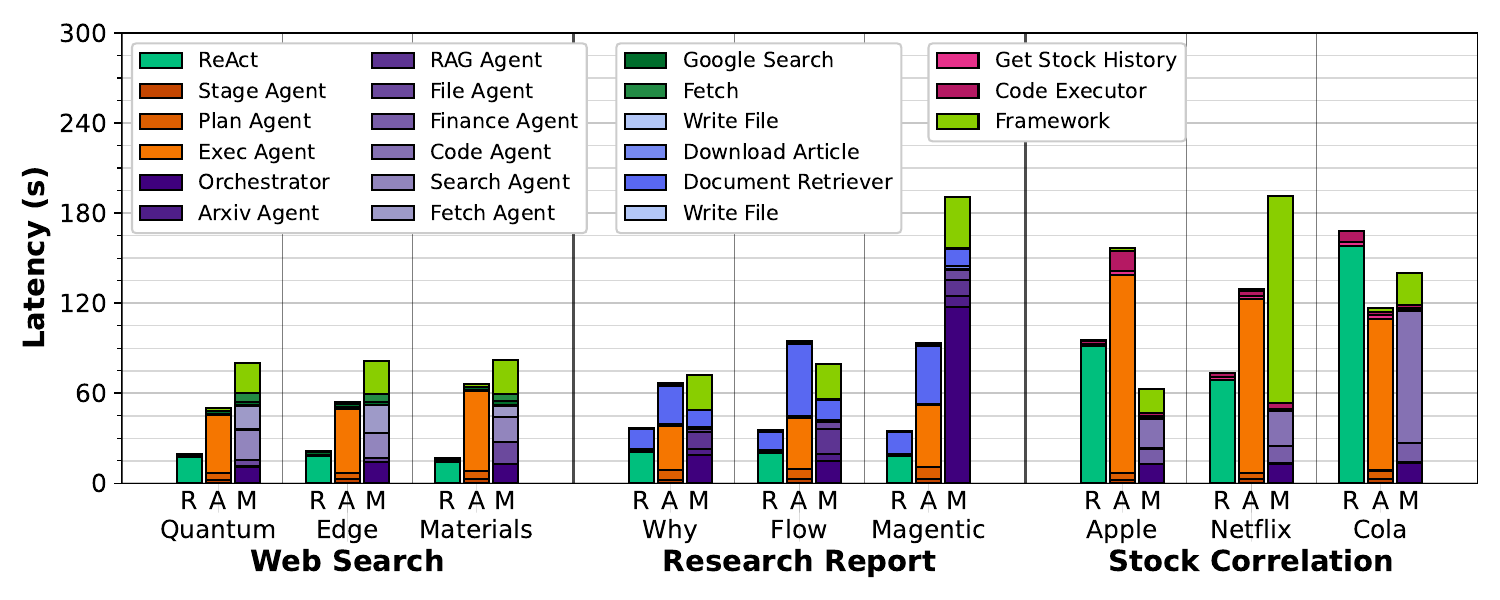}
\caption{Average latency to complete the application across 5 runs for FaaS runs}
\label{fig:Latency-FaaS}
\end{figure}

\agentx in general is slower than ReAct. For example, for the web search application in the `Materials' instance AgentX takes 81.8s while ReAct takes 43.5s on an average over the five runs. This can be attributed to ReAct's direct execution model. Unlike Magentic-One and AgentX, which delegate tasks to specialized agents, ReAct consists of a single agent for all executions. Therefore, latency from additional inferences are absent.  

\agentx in general is faster than Magentic One. For example, for the `flow' and `magentic' instances of research report application, \agentx is faster than Magentic One by 40.1\% and 31.9\%, respectively. This is because Magentic one suffers from additional framework overheads, with a mean framework latency of 30.1s, compared to AgentX and ReAct that have mean framework latencies of 2.0s and 0.1s, respectively, for the web search application across. We believe this is primarily due to the underlying Autogen framework and the network overhead of using AgentOps as the observability tool. However, if the framework latency is ignored, AgentX and Magentic One have similar latency values. 

\agentx is slower than Magentic One for the stock correlation application by an average of 164\%  across all instances. This is because the Magentic One truncates the stock data which results in lower LLM inference time while \agentx has comprehensive data handling where it uses the whole data obtained from Yahoo Finance to plot and format it for better representation, which leads to higher latency but also greater fidelity. 

\begin{figure}[t]
\centering
 	\includegraphics[width=0.5\columnwidth]{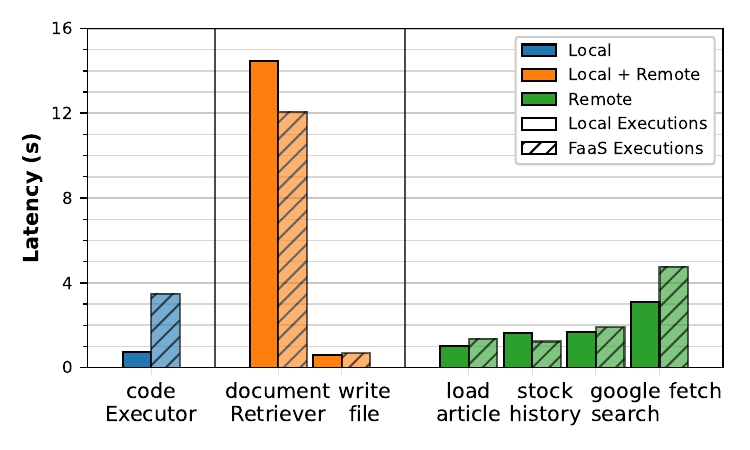}
 	\caption{Tool execution latency comparison across local and FaaS}
    \label{fig:Latency-Tool}
\end{figure}

The latency contribution from LLM invocations and tool invocations depends on the application being executed. Applications that involve quick external API calls have majority of the latency contribution made by LLM inferencing where as applications that involve memory or computationally intensive tools have significant latency contribution made from tool invocations. Tools like `Google Search' and `Get Stock History' have an average latency of 1.7s and 1.6s where as tools like `Document Retriever' has an average latency of 14.1s as shown in Fig.~\ref{fig:Latency-Tool}.

\begin{figure}[t]
\centering
 	\includegraphics[width=\columnwidth]{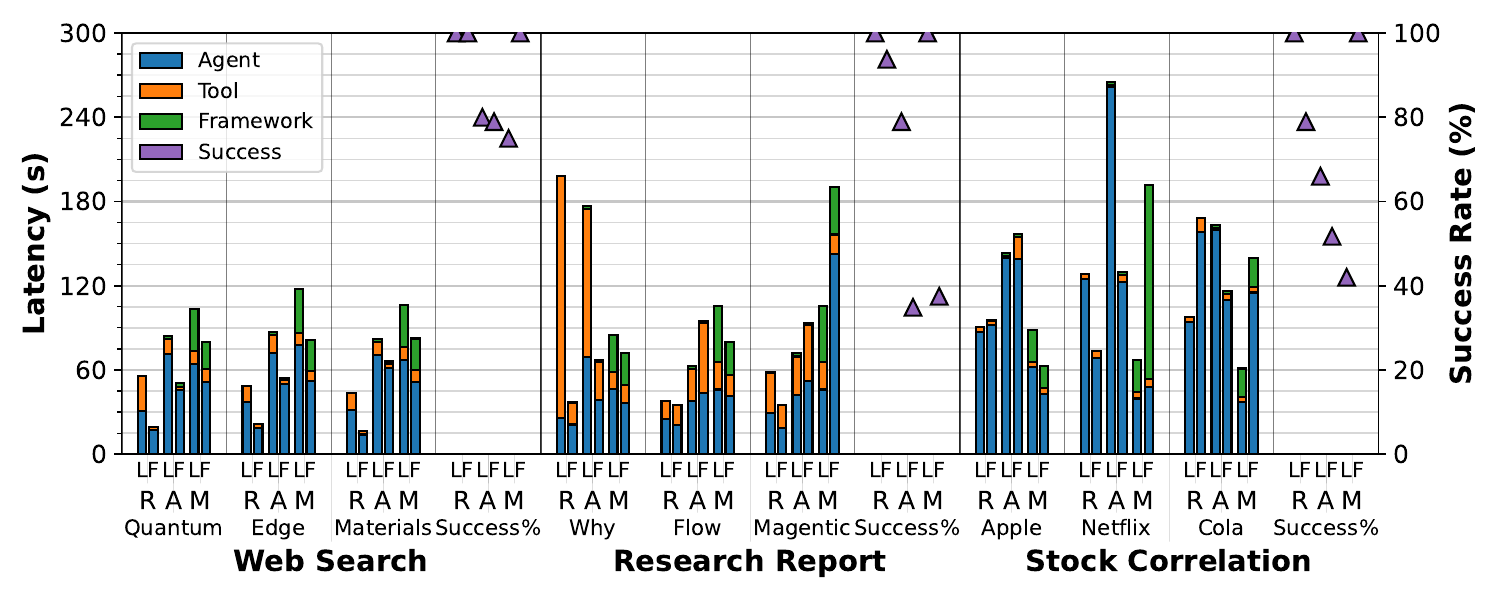}
 	\caption{Overall latency \textit{(bar $\blacksquare$, left axis)} and success rate \textit{(marker $\blacktriangle$, right axis)} comparison between local and FaaS executions.}
    \label{fig:Latency-Faas-vs-Local}
\end{figure}

Latency contribution made by LLM invocation for \agentx for the web search and stock correlation application is 84.4\% and 98.1\% of the total latency where as it is 47.9\% of the total latency for the research report application across all instances. Latency contribution made by tool invocations for \agentx for the web search and stock correlation application are 13.3\% and 0.01\% of the total latency where as it is 49.8\% of the total latency for the research report application across all instances as shown in the Fig.\ref{fig:Latency-Faas-vs-Local}. ReAct and AgentX for the `Why' instance of the research report application has the average total tool latency of 172s and 105s, caused due to the `Document Retriever' tool exhibiting variable latency of 0.77--795s, which causes the total latency to reach high values for some of the runs and the average total latency grows to 176s and 197s for ReAct and AgentX, respectively.

\paragraph{FaaS MCP}
In the serverless MCP setup reported in Fig.~\ref{fig:Latency-FaaS}, the latency patterns largely hold, which includes AgentX being slower than ReAct and faster than Magentic One. For example, AgentX is slower than ReAct by 296\% and faster than Magentic One by 19.9\% for the web search application on average across all instances. This trend is also observed for the Research Report if the Framework latency and the variability of the Document Retriever tool is ignored. 

Magentic One for the Stock Correlation application truncates the stock data during code execution causing faster execution at the cost of accuracy of the output. The average framework latency for Magentic one for the FaaS runs is around 15s on average across all the instances of all the applications. However, due to an outlier, where the framework latency is 613s there is a huge spike for the `Netflix' Instance of the Stock Correlation application in the framework latency. This we believe is due to a network connection issue with the AgentOps observability tool. 

In Research Report application, AgentX is slower for ``Flow'' instance than Magentic-One due an outlier caused by the Document Retriever tool in RAG MCP server execution time, where usually they range from 6--12s, but in run one of AgentX the execution time shot up to 133s that increased the average total latency to 94.5s. 

Additionally, in the ``Magentic'' instance, Magentic-One's average latency rises as the usual latencies ranges from 68--113s but in run 1 of that instance, the latency reached 553s. This is due to multiple retries due to the agents being unable to find and download the right paper from ArXiv and saving it to wrong S3 location. Hence, there is a need to build agentic frameworks with cloud executions in mind, as these frameworks do not deal with read/writes to an external service on a local machine where the paths for inputs/outputs is not strictly enforced.

\paragraph{Comparing Local vs FaaS}\label{sec:results:latency:local-faas}
A direct comparison between FaaS and local execution latency in Fig.~\ref{fig:Latency-Faas-vs-Local} highlights the variability in performance of the FaaS environment. \textbf{L} indicates local MCP server while \textbf{F} indicates FaaS hosted MCP server. The figure includes a high level latency break down into those caused by \textit{LLM inferencing} colored by blue, \textit{tool invocations} in orange, and the remaining \textit{framework} overheads in green. The figure also reports the Success \% of the three agentic patterns, marked by purple triangles on the right Y axis.

While the only difference between the Local and FaaS runs are the deployment of the MCP servers, due to a difference in the tool descriptions present between Local and FaaS, the patterns were not able to utilize fetch tools in FaaS MCP servers. Therefore, a direct comparison between Local and FaaS for the Web Search application is not necessarily fair. However, the key focus of our analysis it the tool latencies between the two deployments of the MCP servers, which can be compared and analyzed. The only completely local tool is the code executor tool which involves the execution of Python code. The tool takes 0.7s on average locally while taking 3.4s on FaaS MCP servers. This increase in latency, we believe, is due to (1) network overhead caused by triggering the Lambda function and returning the output, and (2) a difference in hardware performance for executing the code on the Lambda function. We note a lower latency for the Document Retriever tool and Stock History using the FaaS MCP server, by 16.9\% and 26.5\% than the local. But this cannot be completely attributed to the FaaS performance since it involves remote execution to external services that have variable latencies. For other tools making remote executions such as load article, Google search and fetch, the FaaS MCP servers perform 27.1\%, 13.5\% and 34.8\% slower compared to their local counterparts.

\paragraph{Success Rate}
The success rate of an agentic pattern for an application is calculated as follows. The patterns are executed until 5 successful runs are obtained for each of the 3 instances of an application. The success rate of a pattern for that application is $3 \times 5 = 15$ divided by total number of runs required by that pattern to obtain those $15$ successful runs. ReAct has the highest success rate across all the patterns having 100\% success rate for all the applications for the local set of experiments. The success rates for the Web Search application is 80\% for AgentX, 100\% for ReAct and 75\% for Magentic One. This can be attributed to ReAct having to maintain only a single context window unlike AgentX and Magentic One where necessary context has to be passed between different context windows belonging to different agents. Also, ReAct has a default recovery system where it tries until it succeeds for a maximum of 25 iterations until it generates the \textit{`Final Answer'}. 

AgentX and Magentic One both suffer from lower success rates for all three applications. For example, for the Stock Correlation application AgentX has a success rate of 66\% while Magentic One has a success Rate of 42\%. These failures are due to multiple reasons, from not passing necessary context, using non-existent tools, lack of a recovery system in AgentX, being stuck in a loop of using invalid parameters for tools, and so on. These failures and anomalies are discussed further in the Discussion section~\ref{sec:discuss}.

\subsubsection{Input Token Analysis}\label{sec:results:input-tok}
\begin{figure}[t]
\centering
 	\includegraphics[width=0.8\columnwidth]{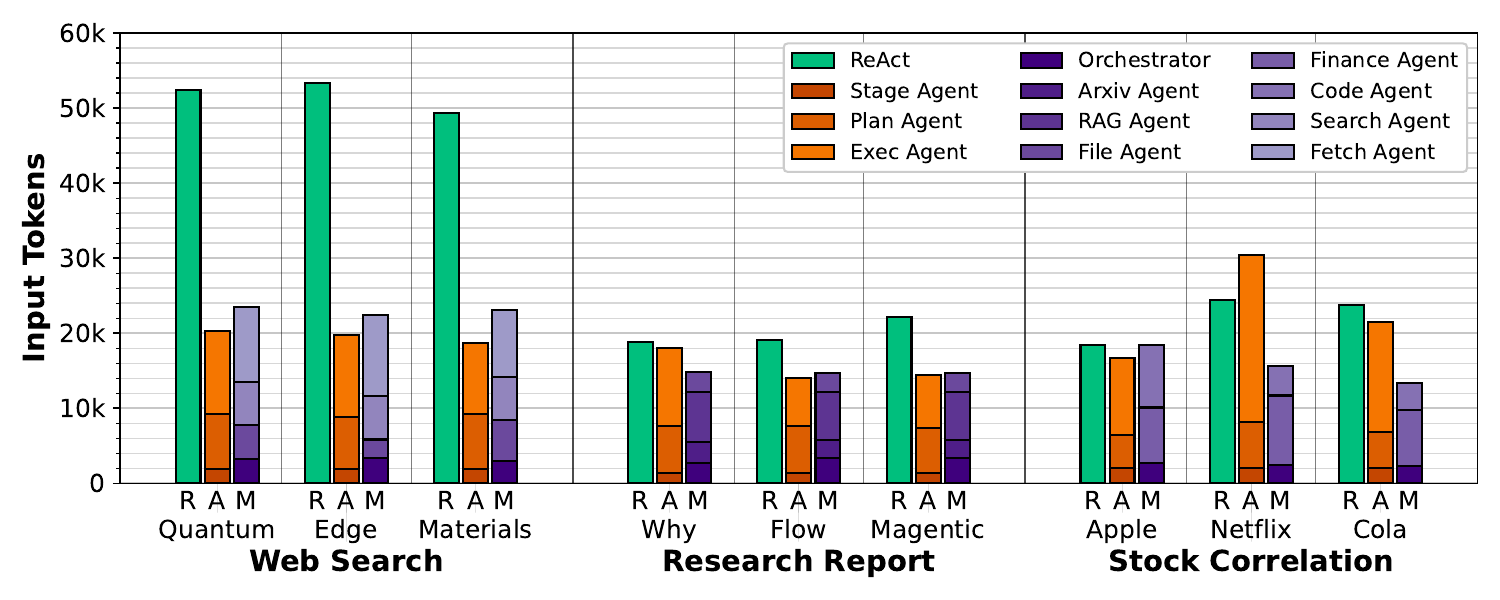}
 	\caption{Average input tokens consumed for the Local executions}
    \label{fig:Input-Tokens-Local}
\end{figure}

The input token consumption values are obtained from the metadata generated by the OpenAI client whenever the model endpoint is triggered. The Fig.~\ref{fig:Input-Tokens-Local} shows the input tokens consumed by the patterns on the Y-axis for the local MCP experiments.

\begin{figure}[t]
\centering
 	\includegraphics[width=0.8\columnwidth]{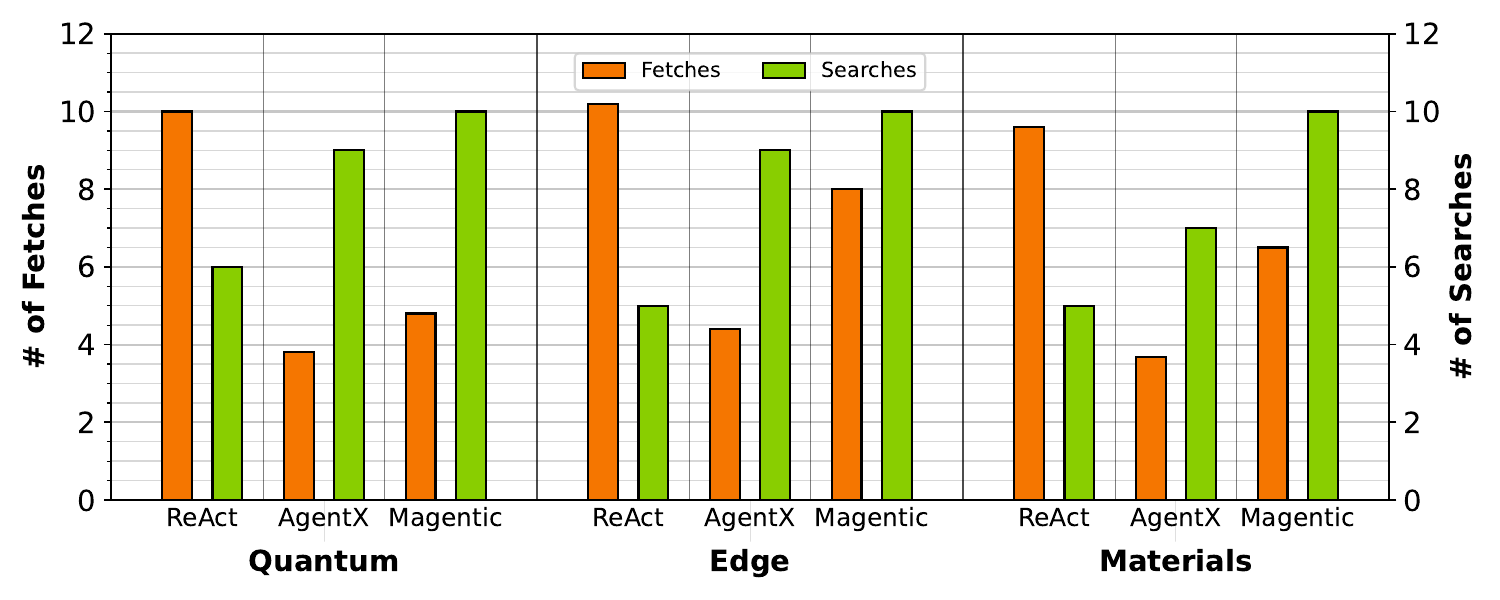}
 	\caption{Number of Fetches and Search Results requested by Agentic Patterns locally}
    \label{fig:fetches}
\end{figure}

In addition, Fig.~\ref{fig:fetches} shows the number of \textit{fetch tool} calls (left Y axis) and the number of search results requested from the \textit{google search} tool (right Y axis), made by the patterns for different instances of the web search application. This is helpful to explain some of the results obtained in the experiments. For context, the google search tool takes two input parameters: query and the number of search results to return. Each search result includes a URL and a small snippet of text from the web page the URL points to. Fetch tool takes the URL of a website and the offset character length as an input and returns clean text content for that web page. It fetches content 5000 characters at a time. So multiple invocations with a higher offset are required used to incrementally get the full content for a large web page.

\paragraph{Local MCP}
In Fig.~\ref{fig:Input-Tokens-Local} we observe that \agentx consistently requires fewer input tokens for the web search application compared to other patterns. It consumes 62.1\% fewer tokens than ReAct and 19.1\% fewer tokens than Magentic One, on average across all instances. ReAct exhibits particularly high input token consumption as it employs a single (long) message history that appends all raw tool and inference outputs directly, which for verbose applications like web search contributes to higher total input tokens. 

Additionally, \agentx triggers fewer \textit{'Fetch Tool'} calls, usually to get the top 3 or top 5 URLs, with an average of 3.9 fetch tool calls, as shown in the Fig.~\ref{fig:fetches}, across all instances of the web search application. In contrast, ReAct and Magentic One perform an average of 9.9 and 6.4 fetch tool calls. This is because \agentx has a dedicated planner agent that creates an optimized execution plan for the application where as ReAct relies on the base LLM's implicit reasoning to guide the execution flow. Magentic One's use of the fetch tool is non-deterministic where it performs 4.8, 8 and 6.5 fetch tool calls on average for the `Quantum', `Edge' and `Materials' instances, respectively. 

The number of tool calls used by the patterns plays a significant role in determining the input tokens consumed. ReAct performing too many fetch tool calls causes it to have a spike in input tokens while \agentx selects and fetches only the top-3 most relevant URL's from all that are returned -- which ranges from 5 to 10 depending on the invocation, with an average of 8.3 URLs in the search result. 

The Google search tool is always used only once by any of the patterns in the entire application workflow execution and it contributes an average of 883 tokens to the total input token consumed, whereas the fetch tool is used a variable number of times by different patterns and it contributes an average of 1063 tokens each time it is being used for the web search application by all the patterns. These token contributions from the tool output add up to the input token consumption.  

For the Research Report and the Stock Correlation applications, the number of tool invocations are similar, as shown in the Fig.~\ref{fig:Tool-local}, and the tool parameters passed on to the tools are also similar based on observations. Therefore, the difference in input tokens contributed from difference in tool usage is negligible. This causes the total input token consumption to be similar to each other. However, due to the non deterministic nature of agentic workflows and certain outliers, there are differences. For example, in the `Netflix' instance of stock correlation \agentx consumes 24.7\% and 94.8\% more input tokens compared to ReAct and Magentic One due to one of the runs where \agentx consistently generates Python code with syntax errors, which caused the execution agent to retry until it succeeded. This led to a high total input token count of 58,101.

Another significant reason for AgentX and Magentic One consuming fewer input tokens than ReAct is the generation of execution results. After AgentX and Magentic One use, say, the Fetch tool, they generate a summary and pass this for later use instead of the entire raw output from the tools. These summaries contribute to fewer tokens downstream. \agentx has a lower input token consumption than Magentic One. For example, for the research report application in the `Why' instance, \agentx has an input token consumption of 14,026 tokens while Magentic One has an input token consumption of 14,674 tokens due to Magentic One passing the fact sheet and the plan as input in every inference it makes. 

With each inference, the entire history of previous inputs and outputs is appended to the subsequent interaction. This causes the input tokens consumed by the LLM to grow across the stages, leading to an expanding context. Tool descriptions also play a significant role in the tokens consumed as they are passed as context in every inference of ReAct, and to the Exec Agent of AgentX. For Magentic One, only the specialized agents will have tool descriptions for the tools they have access to.

\begin{figure}[t]
\centering
 	\includegraphics[width=0.8\columnwidth]{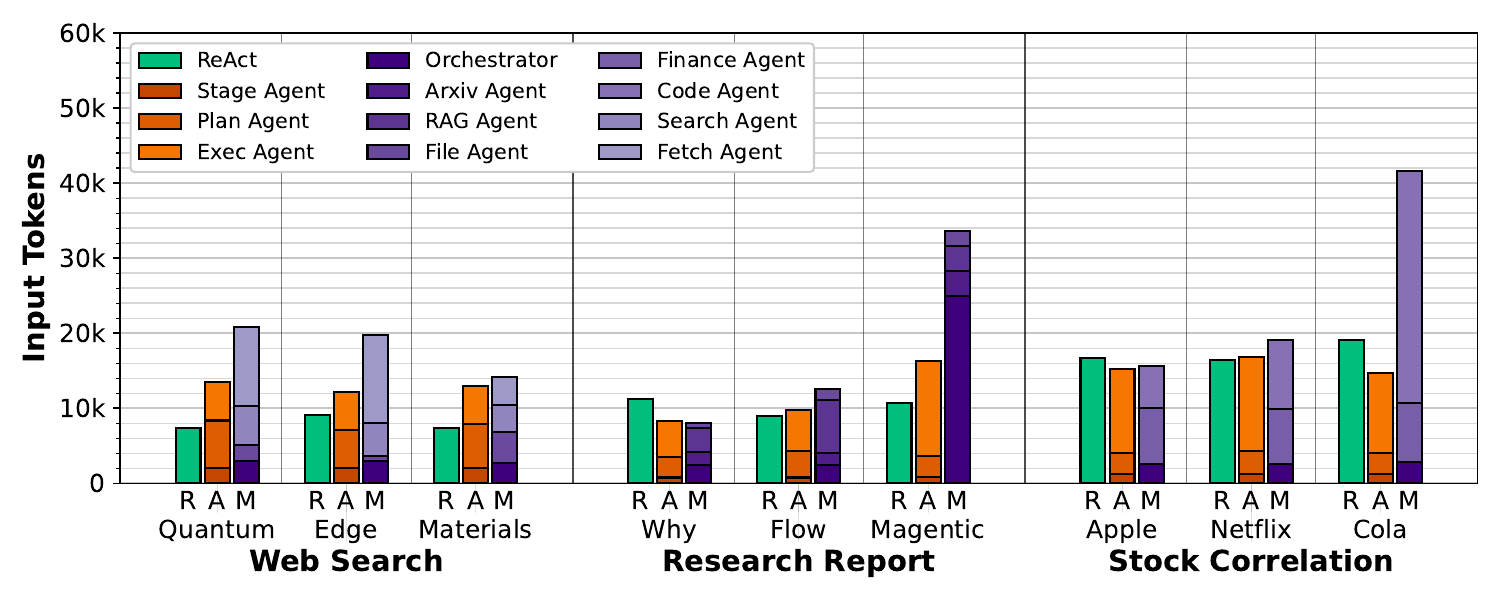}
 	\caption{Average input tokens consumed for the FaaS executions}
    \label{fig:Input-Tokens-FaaS}
\end{figure}

\paragraph{Comparing FaaS vs Local}
Fig.~\ref{fig:Input-Tokens-FaaS} shows the input tokens consumed for the FaaS set of experiments. The total input tokens consumed in the FaaS experiments is lower for all the instances across all the patterns when compared to the local experiments. For example, for the stock correlation application in the `apple' instance, the total input tokens consumed by local executions by ReAct, AgentX and Magentic One are 10.1\% , 9.6\% , and 17.5\% greater compared to FaaS executions. A key reason for this that the FaaS MCP servers only expose a subset of the tools relevant to an application while the local MCP servers list all the tools, as mentioned in the setup section. This causes a decrease in the input tokens consumed by the FaaS experiments, which is not a fair comparison. For example, in \agentx, for the local set of experiments the tool descriptions for the web search, research report and stock correlation applications amount to 1625, 1114 and 1784 tokens for Local MCP, whereas for the FaaS MCP they come to 1459, 714 and 475 tokens. There is also a large variation in the input tokens consumed for the web search application as patterns do not use the \textit{Fetch tool} in the FaaS experiments due to the difference in tool description compared to Local, as mentioned in the setup. However, Magentic One does end up using the Fetch tool due the additional Fetch Agent description provided, which causes it to use more input tokens than ReAct and AgentX, and also being comparable to its Local MCP experiments. For example, the patterns ReAct, AgentX and Magentic One take $54,688$, $22,488$ and $25,194$ tokens for Local MCP while it takes $9,821$, $14,370$ and $22,061$ tokens for the FaaS executions, for the `Edge' instance of the web search application. 

While the total input tokens consumed by the patterns for the FaaS set of experiments are generally lower compared the local set of experiments. However, Ideally there should be little to no difference between the experiments since the input token consumed by an agentic workflow would generally be dependent on workflow execution and not where the MCP servers are deployed.

\subsubsection{Output Token Analysis}\label{sec:results:output-tok}

\begin{figure}[t]
\centering
 	\includegraphics[width=0.8\columnwidth]{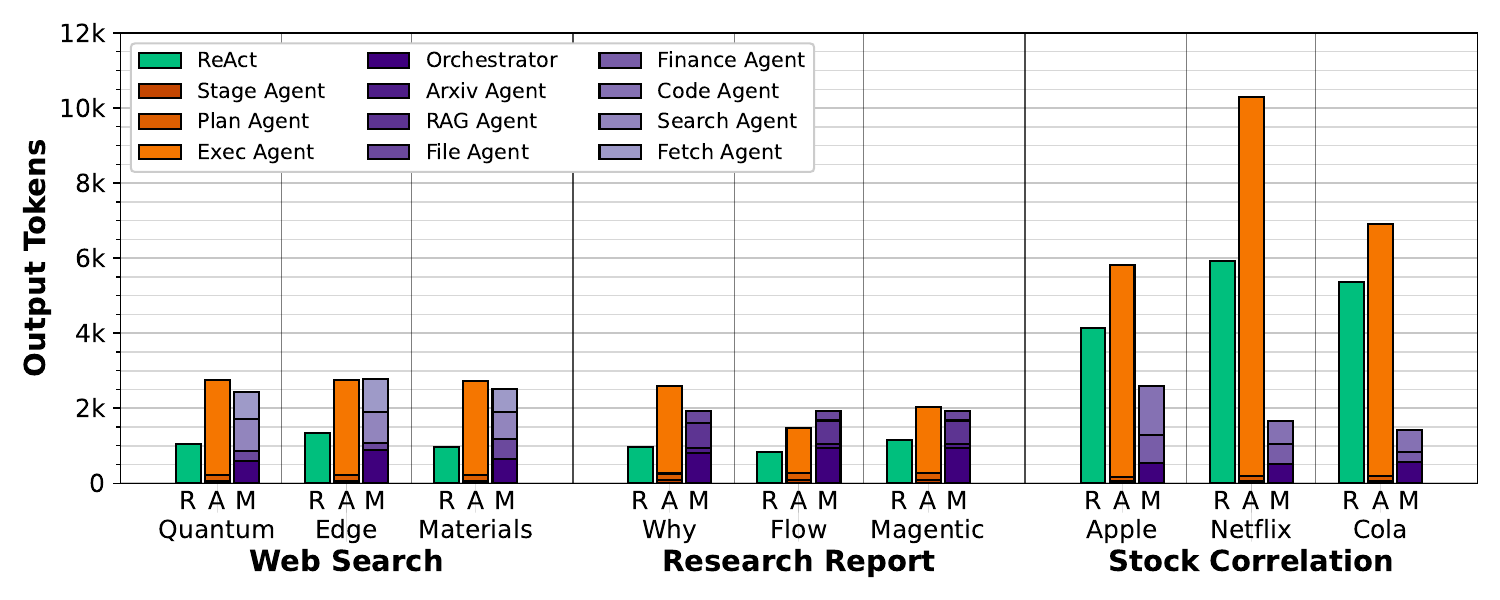}
 	\caption{Average output tokens Generated for the Local executions}
    \label{fig:Output-Tokens-Local}
\end{figure}

\begin{figure}[t]
\centering
 	\includegraphics[width=0.8\columnwidth]{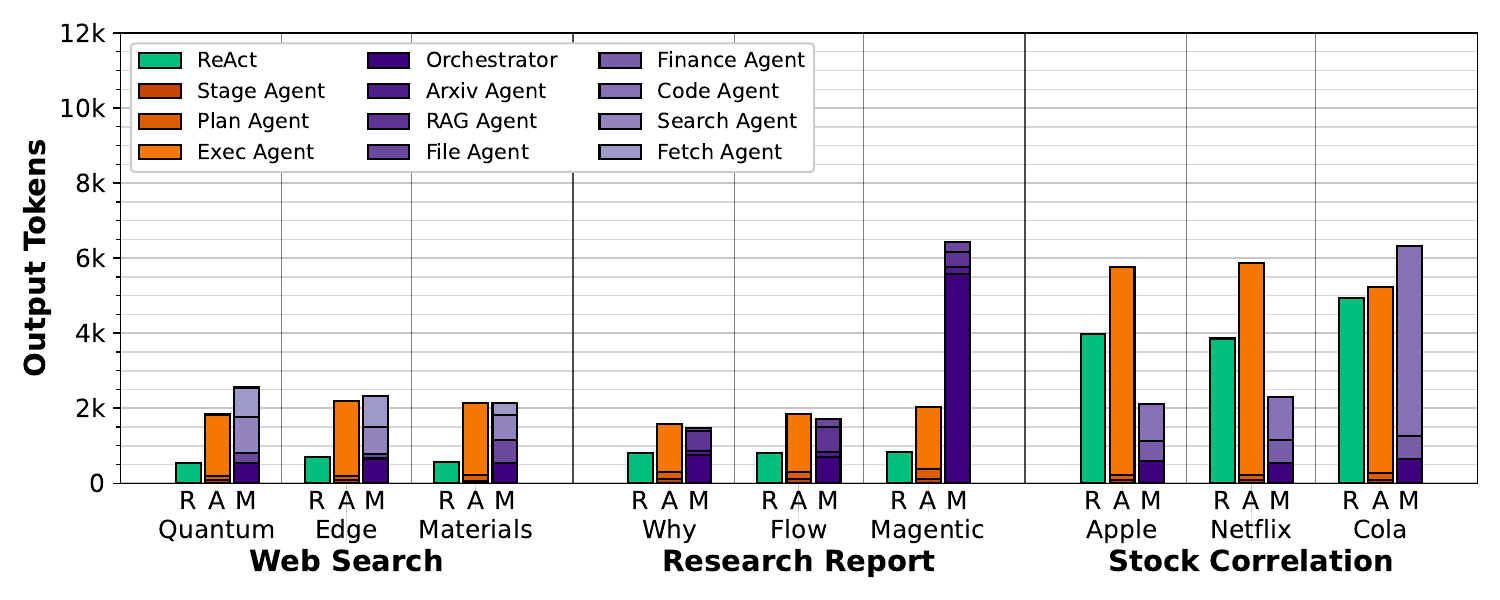}
 	\caption{Average output tokens Generated for the FaaS executions}
    \label{fig:Output-Tokens-FaaS}
\end{figure}

The Fig.~\ref{fig:Output-Tokens-Local} and Fig.~\ref{fig:Output-Tokens-FaaS} shows the output tokens generated by the patterns on the Y-axis, for the Local and FaaS experiments. 
As before, the output token counts are returned in the metadata generated by the OpenAI client when the model endpoint is invoked. ReAct generates fewer output tokens compared to AgentX and Magentic One. For example, for the `Magentic' instance in the Research Report application ReAct, AgentX and Magentic One generates $1149$, $2046$ and $1926$ output tokens on average. This is because ReAct does not generate dedicated execution summaries, unlike AgentX and Magentic One. AgentX generates a summary after every stage that contains the necessary context that needs to be passed on to the next stage, which causes it to generate more output tokens than ReAct. For example, in the web search application, AgentX in one of the runs, in its second stage uses the fetch tool to collect web content which contributes to $4328$ tokens. However, At the end of this stage, the execution agent generates a concise summary of the findings, which amounts to $429$ tokens. This summary generated is then passed to the next stage instead of the original verbose content. This mechanism increases the output tokens generated in the current stage while lowering the input token count for all the subsequent LLM inferences.

Magentic One also has specialized agents that generate execution results after usage of tools. However, the execution results generated by Magentic One are similar to the outputs of the tools themselves, in terms of the length tokens. For example, when the search agent in Magentic One uses the search tool and gathers search results, the execution results largely reproduce the original search results with minimal summarization. Additionally, Magentic One takes two additional inferences to create a fact sheet and a plan of execution. 

The stage and planning agent in \agentx contribute to negligible output tokens. For, example for the `Edge' instance of the web search application, the stage agent contributes an average of $56.7$ output tokens while the planning agent contributes $164.8$ tokens, with the total being $2751.8$ output tokens. This is because the stages and plans generated are not more than a few lines of text.

While this should indicate that \agentx should perform better in terms of output token generation, the creation of additional irrelevant stages mitigates this benefit, and Magentic One generates a similar number of output tokens as \agentx. For example, for the `Materials' instance of the web search application, \agentx and Magentic One generates $2722$ and $2524$ average total output tokens. This is further described in the Discussion, Section~\ref{sec:discuss}. 

For the Stock Correlation application, Magentic One truncates the stock data in its execution results. This causes the code agent to generate a plot using truncated data, or use dummy data when the input data is entirely missing. This causes Magentic One to generate fewer total output tokens than AgentX. For example, AgentX generates $5.18\times$ more tokens than Magentic One's erroneous output for the `Netflix' instance. 

\paragraph{FaaS vs Local}
The output token counts generated by the local and FaaS experiments are largely comparable, with most differences falling within a narrow margin. The minor fluctuations observed are primarily attributed to the inherent non-determinism of LLM's outputs and some outliers in the experimental runs. For example, in the ‘Apple’ instance of the Stock Correlation application, the token outputs for the local frameworks slightly exceeded their FaaS counterparts: ReAct by 4\% and AgentX by 1\%.

The `Cola' instance of Stock Correlation for Magentic One from FaaS (Fig.~\ref{fig:Output-Tokens-FaaS}) generates $6313$ average output tokens while the `Netflix' instance of AgentX on Local (Fig.~\ref{fig:Output-Tokens-Local}) generates $10300$ tokens. These are due to outliers where the generated Python code has invalid syntax. With the recovery system of Magentic One, new fact sheet and plans are created when a specialized agent is unable to execute a task, and the task is attempted to execute again with additional feedback provided. Similarly, the execution agent of \agentx retries executing with updated code until it succeeds. This causes the output tokens to spike to a total of $18,855$ and $20,158$ for these, which causes the overall average to increase. In the `Magentic' instance of the research report application, Magentic One generates $6438$ output tokens on average due to an anomaly where the Arxiv Agent tries to repeatedly use the document retriever tool and also provides an incorrect file path using '\textbackslash' instead of '\textbackslash\textbackslash'. This causes the orchestrator to recreate the fact sheet and the plan multiple times, leading to a higher overall output token of $25,792$.

\subsubsection{Cost Analysis}\label{sec:results:cost}

The cost for an LLM inference is calculated as a linear combination of the input and output tokens, as per the OpenAI API documentation for GPT 4o-mini model (Equation~\ref{eqn:cost}). $\mathcal{C}$ is cost (in USD), $\tau_{\text{in}}$ and $\tau_{\text{out}}$ are the number of input and output tokens, and the cost per 1M input and 1M output tokens are \$~0.15 and \$~0.60 
for the GPT 4o-mini model.

\begin{eqnarray}
\mathcal{C} 
= \frac{\tau_{\text{in}} \cdot \$~0.15 + \tau_{\text{out}} \cdot \$~0.60}{10^{6}}
\label{eqn:cost}
\end{eqnarray}

The cloud cost is calculated for the AWS Lambda FaaS invocations are based on the official pricing~\cite{aws-lambda-pricing}, calculated based on the GB-seconds of memory allocated to the function and the invocation duration, multiplied by the cost-factor of $\$~0.0000166667$ for the ap-south-1 (Asia Pacific - Mumbai) region where all our deployments were done.

\begin{equation}
\text{Lambda Cost (USD)} 
= \frac{\text{Execution Time (secs)} \times \text{Memory Allocated (GB)} 
  \times \$16.6667}{10^{6}}
\end{equation}

\begin{figure}[t]
\centering
 	\includegraphics[width=0.8\columnwidth]{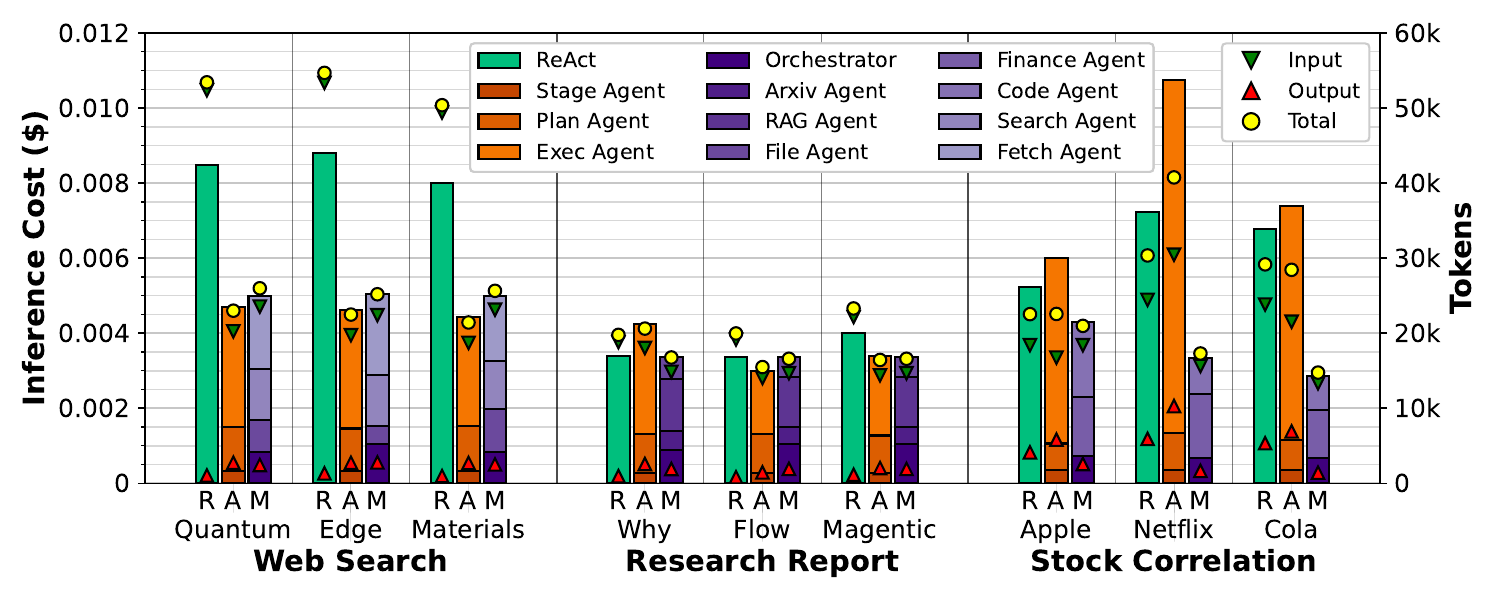}
 	\caption{Cost from LLM invocations from Local Experiments}
     \label{fig:local-cost}
\end{figure}

\begin{figure}[t]
\centering
 	\includegraphics[width=0.8\columnwidth]{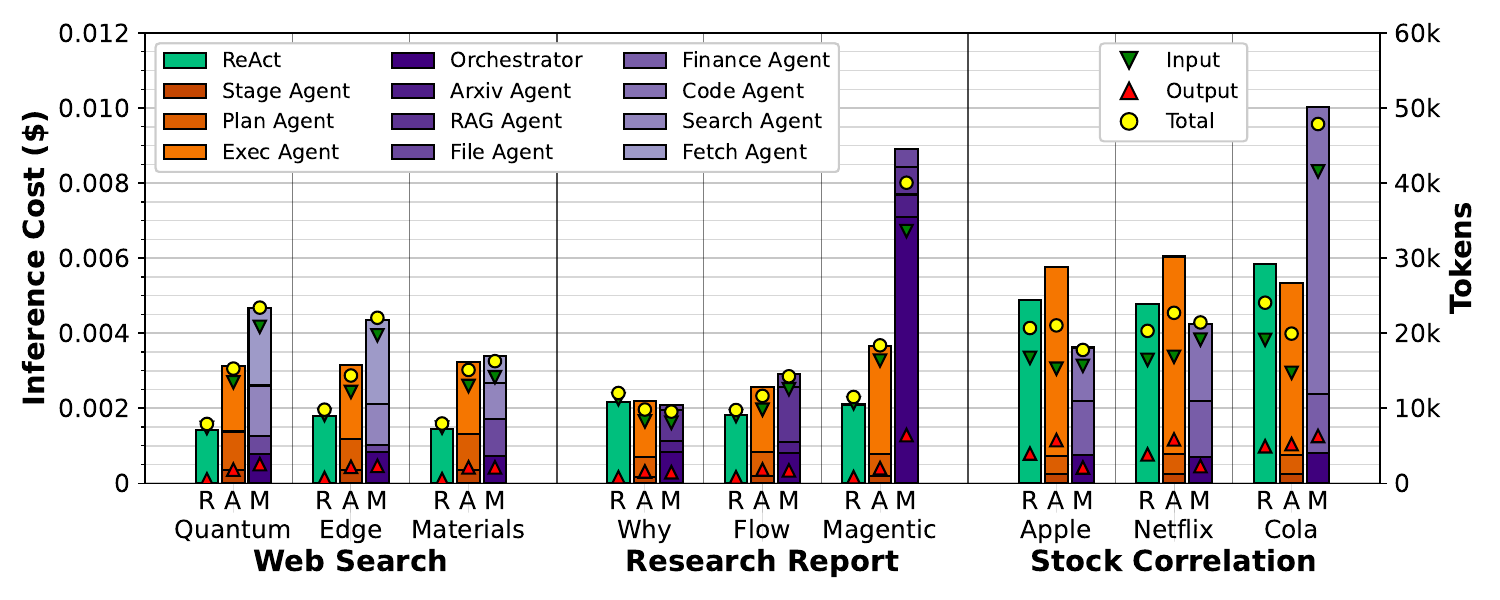}
 	\caption{Cost from LLM invocations from FaaS Experiments}
     \label{fig:Faas-cost}
\end{figure}

\begin{figure}[t]
\centering
 	\includegraphics[width=0.8\columnwidth]{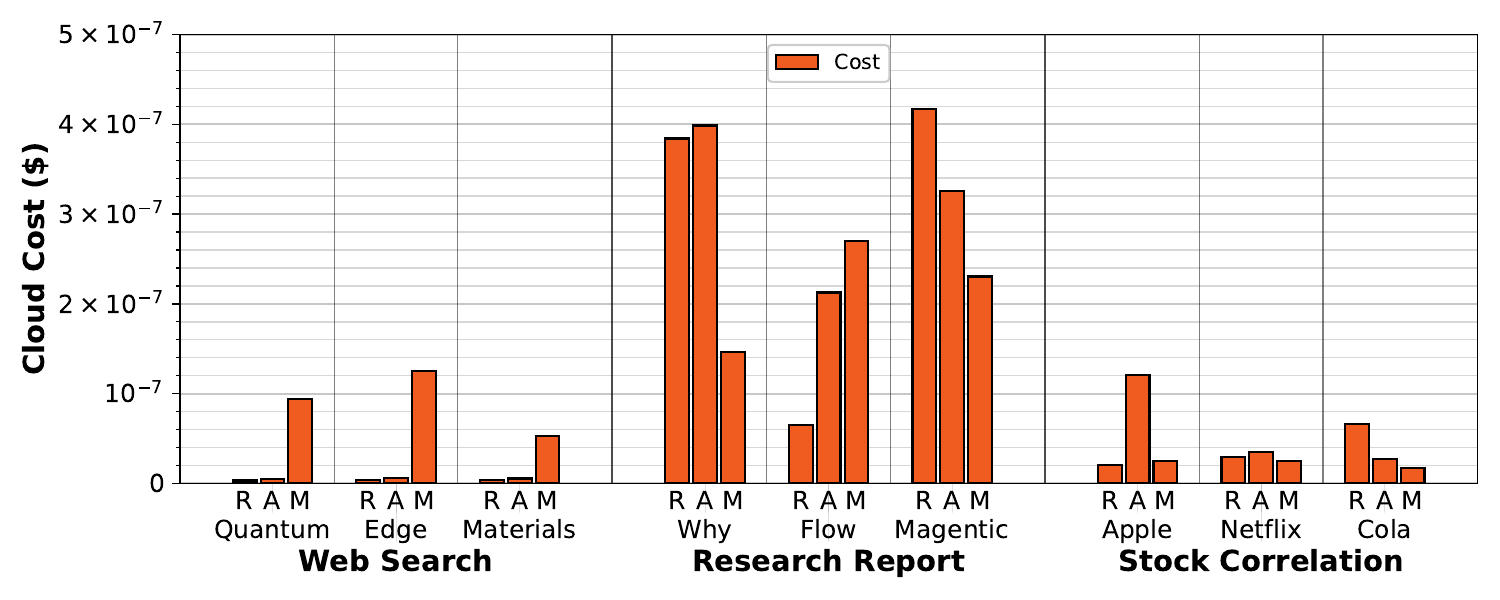}
 	\caption{Cloud Cost obtained from the FaaS Experiments}
     \label{fig:Faas-cloud-cost}
\end{figure}

The Fig.~\ref{fig:local-cost} and Fig.~\ref{fig:Faas-cost} shows the cost of LLM inferencing in US\$ for the Local and FaaS experiments. 
Fig.~\ref{fig:Faas-cloud-cost} shows the cloud cost (US\$) of running FaaS experiments due to MCP server invocations when hosted as a AWS Lambda function. The sum of LLM invocation and FaaS costs give the total cost for FaaS experiments.

\paragraph{Local MCP}
From the Fig.~\ref{fig:local-cost} we can observer that for the web search application AgentX has lower total cost compared to ReAct and Magentic One. \agentx costs 45.7\% and 8.5\% lower cost than ReAct and Magentic One on average across all instances of the web search application. This is primarily due to both ReAct and Magentic One consuming more input tokens, as explained in the input token analysis section (Fig.~\ref{fig:Input-Tokens-Local}). Both AgentX and Magentic One perform better than ReAct for the web search application, primarily due to React using the fetch tool excessively, causing a spike in the input tokens consumed and therefore causing an increase in cost. AgentX does slightly better than Magentic One for the web search application since the latter consumes slightly more input tokens (Fig.~\ref{fig:Input-Tokens-Local}). While the patterns generate different numbers of output tokens for the three instances (Fig.~\ref{fig:Output-Tokens-Local}), these values are too small to significantly affect the total cost of inference.

The patterns produce similar cost values for across all instances of the research report application. For example, the LLM costs for the `Flow' instance for ReAct, AgentX and Magentic One are \$0.013, \$0.012 and \$0.013, respectively, which are comparable because the input and the output tokens for these patterns are similar (Fig.~\ref{fig:Input-Tokens-Local}). For the stock correlation application, \agentx is costlier than both ReAct and Magentic One. For example for the `Apple' instance AgentX is 14.6\% costlier than ReAct and 39.4\% costlier than Magentic One. The higher cost than Magentic One is due to the erroneous truncated the stock data output of Magentic One, leading to lower token usage. AgentX is costlier than ReAct due to the generation of execution summaries, which for applications like stock correlation does not have much scope of reduction and leads to unnecessary costs.

\paragraph{Comparing FaaS versus Local}
A comparison between the FaaS and the local runs demonstrates similar findings from that of the input and output analysis. For the web application, the cost of ReAct in FaaS is much lower than \agentx and Magentic One since it never used Fetch tool, unlike its local counterpart, where it excessively used the fetch tool. For example, in the FaaS experiments, for the `Edge' instance of web search, ReAct costs 46.1\% less than AgentX and 58.8\% less than Magentic One where as in the Local experiments ReAct costs 90.9\% more than AgentX and 74.9\% more than Magentic One. Additionally, outliers in instances causes a spike in the average total cost. For example, the outlier in `Magentic' instance in research report application by Magentic One causes the total cost to spike to \$~0.13 for FaaS.

The cost due to Lambda function executions are negligible compared to the cost incurred by LLM inferencing (Fig.~\ref{fig:Faas-cloud-cost}), by 2 orders of magnitude. For example, the LLM inferencing cost of AgentX for the `Quantum' instance of the web search application is \$~0.012 while the cost contribution from triggering the MCP servers deployed as Lambda functions is just \$~0.0000000044. There is also direct correlation between the tool invocations (Fig.~\ref{fig:Faas-Tool-local}) and the Lambda cost, as expected.

\subsubsection{Tool Invocations Analysis}\label{sec:results:tools}

\begin{figure}[t]
\centering
 	\includegraphics[width=1\columnwidth]{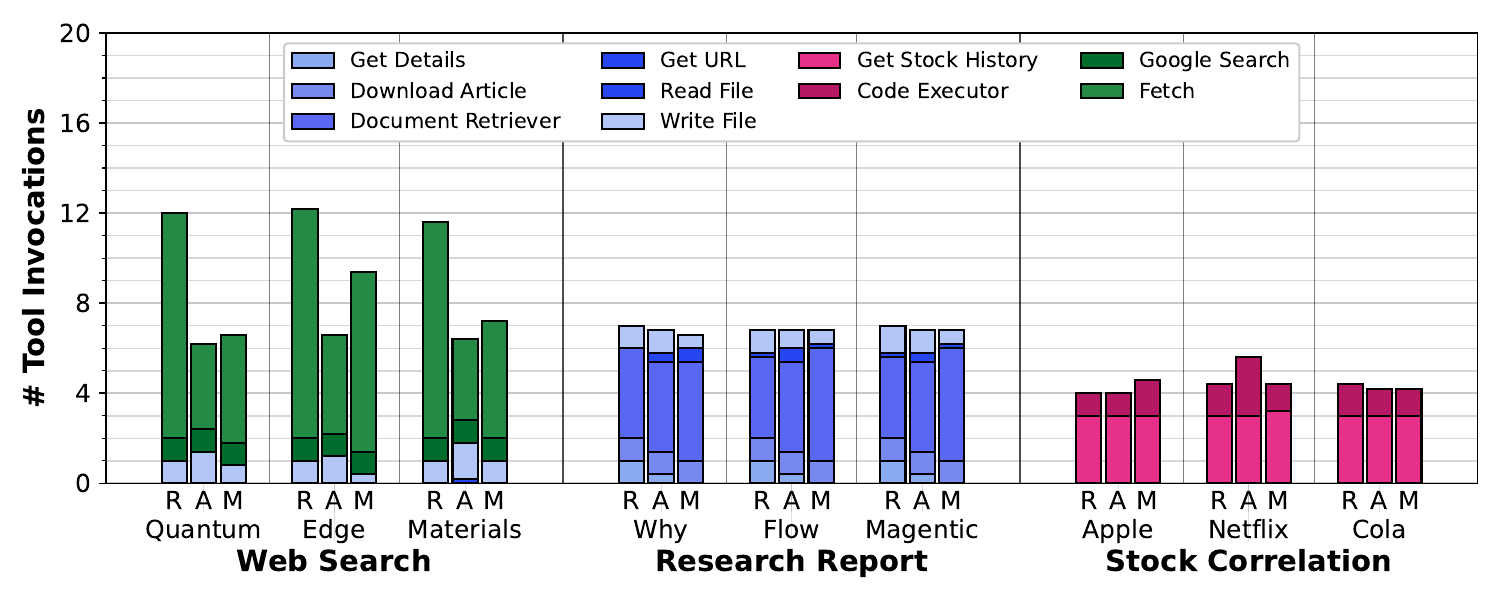}
 	\caption{Average tool invocations for the local experiments}
     \label{fig:Tool-local}
\end{figure}

\begin{figure}[t]
\centering
 	\includegraphics[width=1\columnwidth]{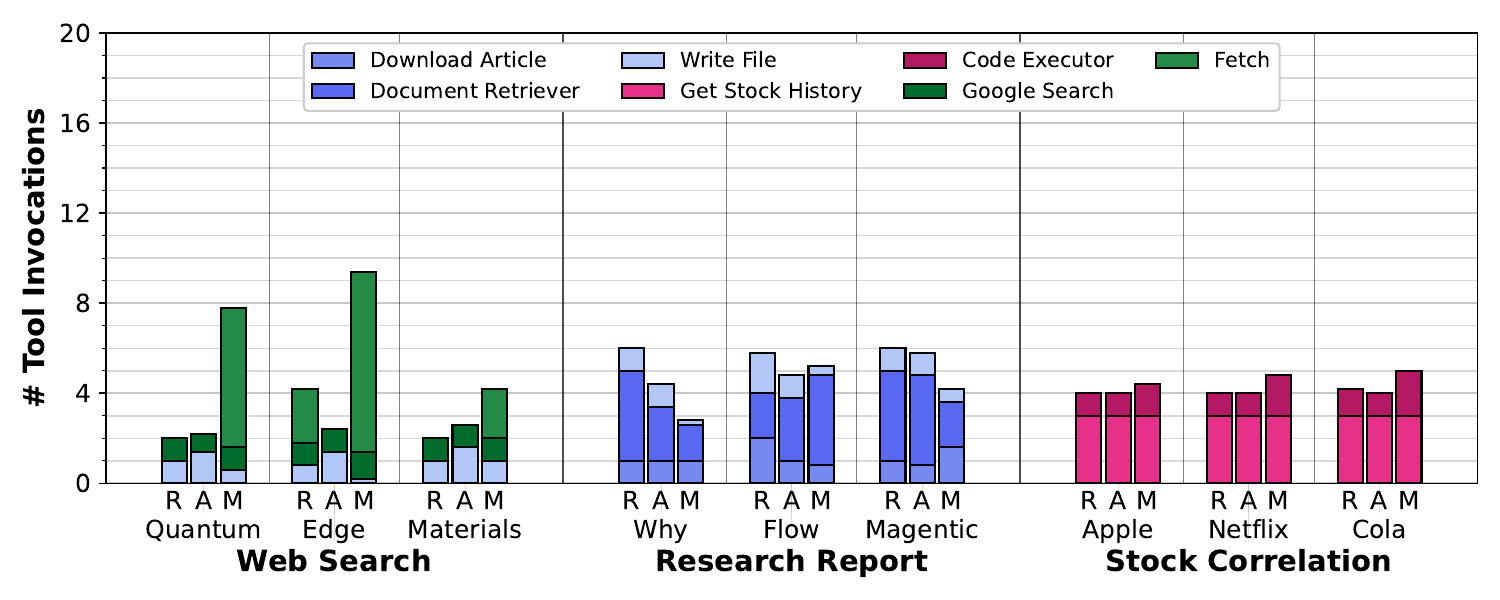}
 	\caption{Average tool invocations for the FaaS experiments}
     \label{fig:Faas-Tool-local}
\end{figure}

Figs.~\ref{fig:Tool-local} and~\ref{fig:Faas-Tool-local} represent the average tool invocations for the local and the FaaS experiments. 
The number of tool invokes adds up the total number of tool calls made by the LLM in the agentic workflow. 
The tool invokes are represented by stacked bars where each stack denotes the contribution of a specific tool to the total number of tool invocations. The tools are also colored such that tools primarily used in an application use shades of the same primary color. Tools used in the web search application have shades of green, tools used in the research report application have shades of blue and finally, tools used in the stock correlation application have shades of red. 

\paragraph{Local MCP}
From Fig.~\ref{fig:Tool-local} we can observe that the tool invocations across the patterns are relatively the same for the stock correlation and the research report application. For example, ReAct, \agentx and Magentic One take and average of 6.8 tool invocations for the `Flow' instance of the research report application. This is because the number of tool calls required to complete the execution of this application is static in nature, i.e., there are not many other ways to complete this application correctly, unlike the web search application where there are alternate pathways to completion to different levels of detail. For the web search application we can see significant variability, due to the variable number of times the fetch tool is used, as described in Fig.~\ref{fig:fetches}. For example, for the `Materials' instance, ReAct, \agentx and Magentic One takes 11.6, 6.4 and 7.2 average tool calls.

\paragraph{Comparison of FaaS vs Local}

For the same `Materials' instance of the web search application, ReAct, \agentx and Magentic One take
2.0, 2.6 and 4.2 average tool invokes for the FaaS set of experiments in Fig.~\ref{fig:Faas-Tool-local}. This contrast is due to the lack of use of the Fetch tool in the FaaS experiments due to the difference in tool descriptions between Local and FaaS. Similarly, for the `Edge' instance ReAct, \agentx and Magentic One make 12.2, 6.6 and 9.4 tool invokes for Local while it takes 4.2, 2.4 and 9.4 tool invokes on FaaS. Magnetic One successfully utilized the Fetch tool during its execution because the Fetch Agent was passed the mandatory additional description provided to the Orchestrator: \textit{``Remote AWS LAMBDA function MCP server for fetching web content in various formats, including HTML, JSON, plain text, and Markdown.''}, giving it the necessary context to correctly delegate the task to the agent, namely, \textit{``to retrieve the relevant content in a suitable format (preferably HTML or plain text) from the search results returned by the SerperAgent.''}

In general, we observe similar results for the stock correlation application. For example, for the `Apple' instance of the stock correlation application ReAct, AgentX and Magentic One takes 4.0, 4.0 and 4.6 average tool invokes locally where as it takes 4.0, 4.0 and 4.4 average tool invokes on FaaS. This is again because the workflow execution tends to be deterministic. The only difference in the tool invokes are observed if the code executor tool fails, which happens when invalid Python code is generated.

\subsubsection{Agent Invocation Analysis}\label{sec:results:agent-inv}

\begin{figure}[t]
\centering
 	\includegraphics[width=0.9\columnwidth]{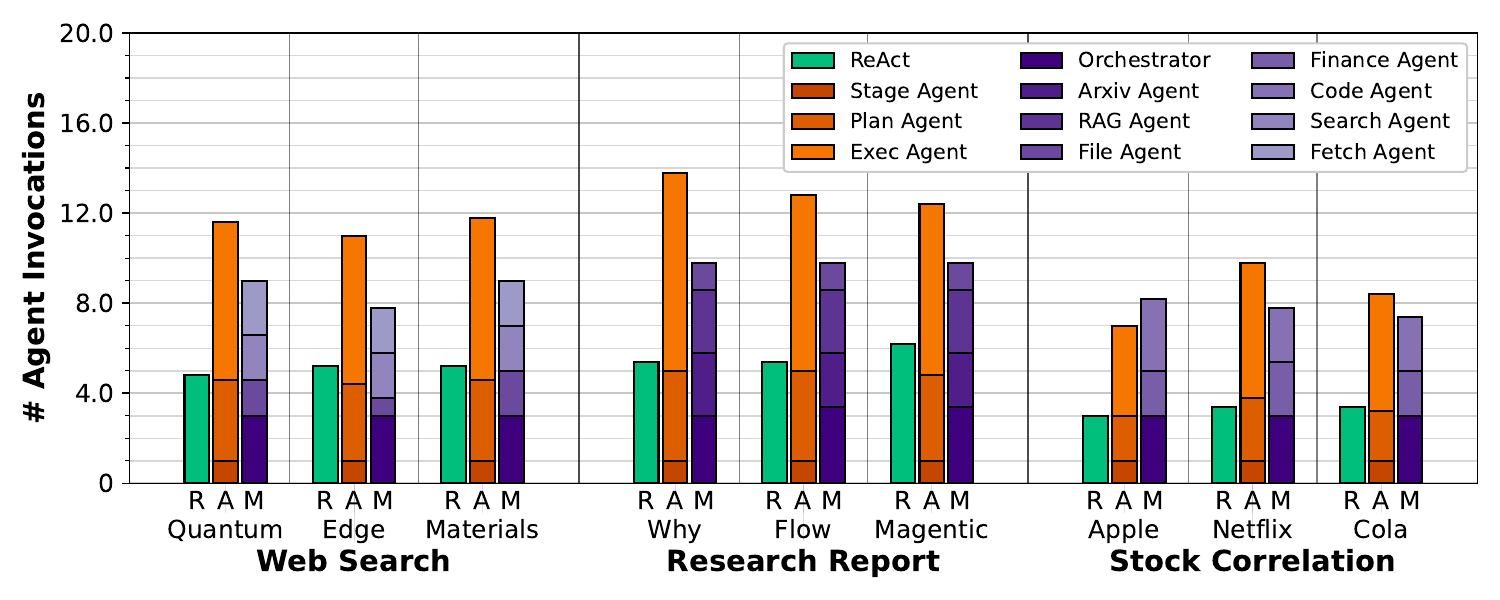}
 	\caption{Average agent invocations for local experiments}
    \label{fig:Agent-local}
\end{figure}

\begin{figure}[t]
\centering
 	\includegraphics[width=0.9\columnwidth]{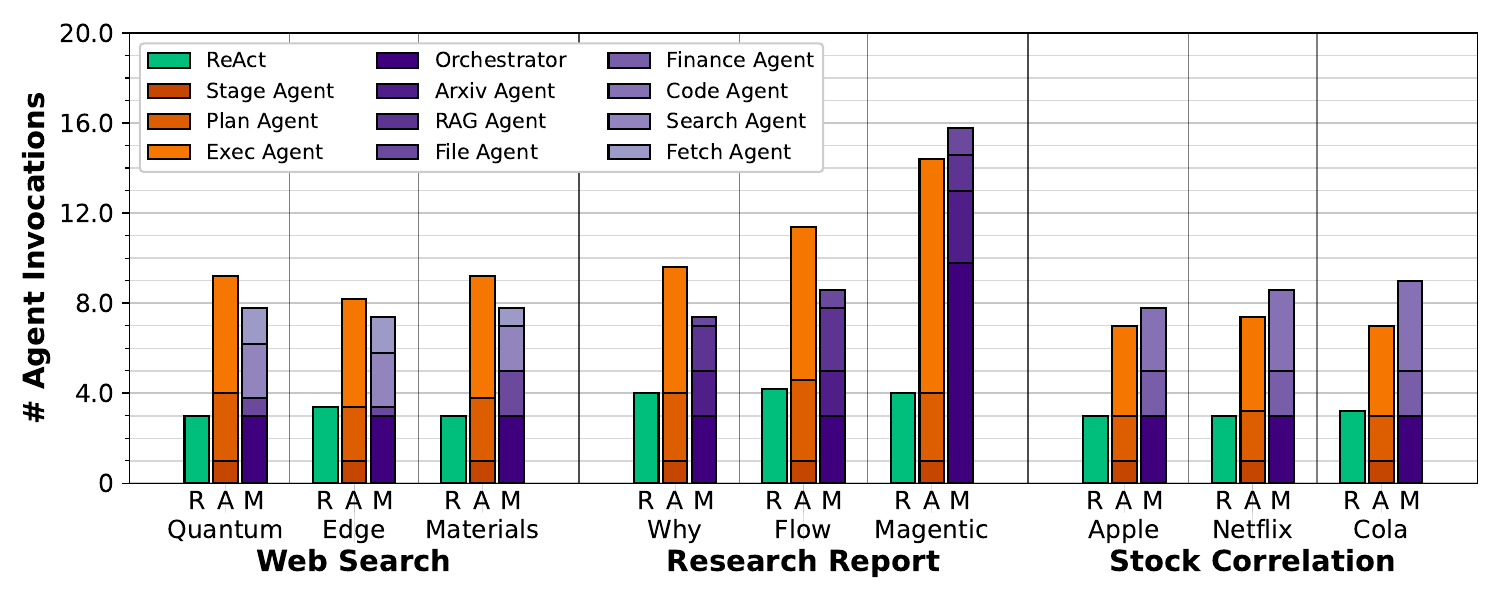}
 	\caption{Average agent invocations for FaaS experiments}
    \label{fig:Agent-Faas}
\end{figure}

Figs.~\ref{fig:Agent-local} and~\ref{fig:Agent-Faas} represent the average agent invocations for the local and the FaaS experiments. 
The number of agent invokes made in a run is sum of the LLM inferences made by each agent in the workflow. The agent invokes are represented by stacked bars where each stack denotes the contribution of a specific agent belonging to a pattern.

\paragraph{Local MCP}
From Fig.~\ref{fig:Agent-local}, ReAct has the fewest agent invocations across all instances of all applications. For example, it takes 55.9\% and 41.2\% fewer inferences than AgentX and Magentic One for the web search application. Unlike AgentX and Magentic One, ReAct uses a single agent. AgentX always uses the Stage agent at the start of the workflow and the planning agent once for every stage. Similarly, Magentic One takes two inferences initially to create the fact sheet and the plan. Additionally, its specialized agents can trigger the Orchestrator once the task assigned to them is complete.

\paragraph{Comparing FaaS vs Local}

Ideally the number of agent invocations should be the same regardless of the MCP server deployment strategy on local or FaaS.
However, from Fig.~\ref{fig:Agent-local} and Fig.~\ref{fig:Agent-Faas}, we see that there are a few variations.

The Agent Invocations for the research report and the stock correlation applications remain similar for Local and FaaS. For example, ReAct, AgentX and Magentic One invoke agents 3.4, 9.8 and 7.8 times locally for the stock correlation application for the 'Netflix' instance, while they make 3.0, 7.4 and 8.6 invocations for its FaaS counterpart. This is because of the predictable nature of this application's execution.

The `Magentic' instance of the research report application for Magentic One has 38\% fewer agent invocations locally compared to FaaS because of the outlier caused where the agent fails to use the document retriever tool as the file path was not properly formatted (\S~\ref{sec:results:output-tok}).

ReAct and AgentX have lower agent invocations on FaaS compared to their local variants due to not using the Fetch tool. For example, ReAct and AgentX perform 5.2 and 11.0 agent invocations for the `Edge' for the local experiments while it took 3.4 and 8.2 agent invocations on FaaS experiments, due to divergence in the tool descriptions; ReAct does not use the Fetch tool on FaaS while AgentX reduces one of its stages. Magentic One had better description of the Fetch tool and therefore has similar tool invokes across local and FaaS.

\section{Discussion\pgs{2}} \label{sec:discuss}
Finally, we discuss several interesting observations, anomalies and insights drawn from our experimental runs.

\subsection{AgentX Anomalies}
AgentX for the \textit{Web Search Application} creates three stages generally: Search the Web with the query, Use the Fetch Tool for the most relevant URLs, and Summarize the contents and write it into a text file. Occasionally, it splits the content summarizing and writing into two separate stages. However, this inclusion of a separate writing stage faces an unexpected problem. The tools passed to the agent during the summarization stage includes the tool to write to a file, which causes the execution agent to write the summarized content into a text file. However, since there is already another stage specifically to write to a file in the next stage, it would write the summarized content to the text file again. This behavior is observed even after explicitly instructing the Stage Agent to combine the summarizing and writing to a file task, which causes unnecessary latency and cost in some of the runs. 

The Planner Agent in the web search application fetches the top three and sometimes the top five URLs out of the available results using the google search tool. The Execution agent chooses the top URLs based on the small text snippet the search results provide. While this approach can reduce the cost of the application compared to fetching all 10 results, the quality of the content generated is affected. Also, \agentx fails to occasionally complete the the web search application by not writing the file at the end, which are considered as failed runs. These are the primary reason for the lower success rate in Fig.~\ref{fig:Latency-Faas-vs-Local}.

\agentx for the \textit{Research Report} application generates the following four stages: Retrieve the article metadata, Download the article, Query the downloaded document for information, and Save the summary as a text file. During the Planning stage, the tool parameters are sometimes not explicitly mentioned, e.g., the file path to the PDF is missing when attempting to query the document using RAG. This causes the execution agent to use dummy values when the necessary context is not available. Such runs leads to failures since there lacks a dedicated recovery system in \agentx. 

\agentx create two stages for the \textit{Stock Correlation} application would: One to get the historical data and another to create the plot. Occasionally, an additional stage to process and consolidate the data is also created. A separate stage to save the PNG is generated even when explicitly asked not to, as it incurs unnecessary cost and latency. The execution results for this application is the entire tool output, which also causes unnecessary increase in cost for \agentx.

\subsection{ReAct Anomalies}
ReAct for the Web Search application directly uses the google search tool and returns exactly five search results, and then fetches from these 5 search results. To understand ReAct agent's behavior, it's essential to first understand the 'fetch' tool. The fetch tool is designed to retrieve text content from a web page and takes the following key parameters: 'url', the web address of the page to retrieve; 'max\_length', an integer that indicates the maximum number of characters to return from the web content which by default is set to 5000 characters; and 'start\_index', an integer specifying the character offset from which to begin fetching, allowing for the retrieval of content in sequential chunks. The tool description includes explanations for these parameters. 

ReAct initially calls the fetch tool on the five URLs it collected from the search tool. However, since the fetch tool has a limit of 5000 characters by default the tool would return partial content of the web page and provides the line \textit{<error>Content truncated. Call the fetch tool with a start\_index of 5000 to get more content.</error>} in the end of the tool output. This increases the total number of fetch tool calls to around 10, as shown in Fig \ref{fig:fetches}. This repeated invocations of the fetch tool on the same URL is not seen in \agentx and Magentic One. 

ReAct for the \textit{Research Report} application follows the intended flow of execution of downloading the Article, using the document retriever, and then writing to the file. However, just like the other patterns, it occasionally uses irrelevant tools like the Get URL and the Get Details tool. ReAct follows the intended flow of execution for the \textit{Stock Application} where it gets the stock data and then plots it. However, generating this plot code takes the most amount of time since the code itself contains the stock data inline. We also observe that ReAct makes frequent code errors compared to other patterns, hinting at potential increase in code generation errors as the context window fills up.

\subsection{Magentic One Execution Flow}
Magentic One starts with the orchestrator answering a survey where it initializes a fact sheet which includes facts given, facts to look up, facts to derive, and educated guesses. The Orchestrator then creates a plan for the task given the fact sheet and the team information.  The Team in Magentic One is created such that it involves multiple agents. The Orchestrator is only exposed to the agents through their agent descriptions. These agents are created such that each MCP server belongs to an agent and the agent descriptions are created keeping in mind the tools exposed to the agent through the MCP server. Unlike for React and \agentx, the Orchestrator of Magentic One also requires additional description of the MCP servers to be provided to it, which helps it during planning, e.g., allowing it to use the Fetch tool in the FaaS experiments.

All the Agents involved in the execution are given the fact sheet and the plan made by the orchestrator. The agents do not send their entire context window to the next agent. Instead, once the agent calls the tools their reflection of the tool outputs are send to the next agent. Once the plan made is successfully completed the orchestrator makes one additional inference to give the final answer to the user.

\subsection{Magentic One Anomalies}
The plan made by the orchestrator for the \textit{Web Search} application involves using the Search agent to get the URLs, then using the Fetch Agent to fetch the contents from those URLs, and then writing the summarized results through the File system agent. The number of fetch tool calls made by Magentic One does not follow a trend. It sometimes fails to write to a file, either never invoking the file system agent or never using the write tool, and instead generates a summary for the user. 

The plan made by the orchestrator for the \textit{Research Report} application involves searching and downloading the paper through the Arxiv Agent, then using the RAG agent to extract relevant sections from the paper, and finally using the file system agent to save the summary into a text file. The plan always includes a verification step to check if the text file has been created or not and the contents of the file are present or not. However this part of the plan never gets executed in any of the runs and the user task is marked completed immediately after writing the summary to the file. 

Three out of the fifteen runs have a distinctive flow of execution, where the Arxiv agent plans to get article details followed by downloading the article. It uses the Get Article details tool but instead of downloading the article, it transfers the control prematurely to the RAG agent. The RAG Agent, without having the file path of the downloaded article, uses a dummy file path thar leads to the tool returning an error output since there is no file present in that path. Since Magentic One has a recovery system, the RAG agent then sends the control back to the Arxiv Agent through the Orchestrator and requests it to download the PDF again. This also triggers two additional inferences, one to update the fact sheet and one to describe the reason for the failure to then create a new plan to overcome. Failure cases involved here includes the File system agent not using the write tool or not transferring the control to the file system agent entirely, just like in the web search application. 

The plan made by the orchestrator for the \textit{Stock Correlation} application involves collecting the stock data through the Yahoo finance agent and then using the code executor agent to generate a plot. The Yahoo finance agent collects the stock data through the Get Stock Data tool. But since only the reflections/summary of the tool outputs are sent to the next agent, the code executor agent occasionally 

just receives the text: ``I have successfully retrieved the data for the stocks'', instead of the actual stock data. The code execution agent then creates a plot with dummy data and simply provides a comment in the code saying ``replace with actual data''. This is a common occurrence in Magentic One and are considered as failed runs. In cases where the Yahoo finance agent generates the stock data in its output, the data is always truncated such that only a small fraction of the original data is preserved. This causes the code execution agent to generate a plot without many data points, reducing the quality of the output. Such runs are still considered as successful runs since the workflow generates a plot; however its accuracy score drops.

\subsection{Pattern Behavior and Insights}

Agentic patterns exhibit significant difference in their execution even when deployed in an identical application environments. This highlights their non-deterministic nature and how architectural design dictates execution strategy. 
While the ideal execution path involves using the tools in a particular sequence to reach the accurate information, we observe that the agentic patterns frequently adopt suboptimal strategies. For example, Magentic One occasionally completes the web search application without making use of the fetch tool. Another interesting behavior is the calling of redundant tools to complete a workflow, which can drive up costs and latency. For example, in the research report application, the patterns make use of the Get URL and Get Details tool even when their outputs have no effect. 

However, we also observe the agentic patterns often follow the ideal execution path even without dedicated planning and reasoning. For example, ReAct uses the Fetch tool twice on the same URL to collect complete data from that website. These behavioral differences suggest that certain agentic frameworks are inherently better suited to specific task types. Efficiently designing robust agentic workflows is critical for their deployment in real-world applications. Without proper guardrails and retry mechanisms, agents can fail, provide incorrect information, or produce improper outputs. For example, we observe that the patterns never using the Write tool even when they have planned to use it. These failures can occur unpredictably, which necessitates rigorous verification systems when operating these agentic workflows at scale, especially for long-horizon tasks. Therefore, guarantees of task completion and output quality remain a challenge.

\subsection{MCP: An Agent's Toolkit}
MCP servers gives agentic frameworks access to to use wide range of tools. However, the problem of selecting a specific tool for a given task still lies with the underlying LLM model based on the tool descriptions and context. Tool descriptions inform the model about the nature of the tool. But due to non-deterministic nature of LLMs combined with sometimes vague tool descriptions, the agent may still fail to use the right tool even when the are explicitly instructed to do so. Additionally, LLMs can also call the tools in the MCP servers with incorrect parameters. This could be resolved by providing an example usage or by using more capable models. This motivates the need for a dedicated LLM inference for tool selection and usage, which can help the workflow execution align to the ideal flow of execution.

MCP servers running in serverless environments like AWS Lambda add their own concerns on top of existing issues. A key challenge is the lack of access to a persistent local storage that spans the session, unlike a local setup. Therefore, read/write of a file has to done solely using an external storage service like S3. Additionally, in the case of code executor server, since Lambda does not allow runtime installation of Python packages, by default, we are required to 
prepackage all dependencies, which bloats the function size and adds unnecessary packages even when not required. Servers which use multi-threading internally do not work in a Lambda environment and workflows which involve scraping data from a website like Yahoo finance may have their requests, which originates from AWS IP address, get blocked due to request requests from this IP range.

MCP servers, by design, are intended to be long running programs, waiting for LLM applications to trigger them. Their main advantage is one of allowing agents of any framework to access the tools using a standard API based on their descriptions. However, for MCP servers to scale, deploying them as FaaS function on cloud is an viable option. However, the cost of this deployment is not trivial to calculate. The type of tools present in the MCP server contribute to the total cloud cost. Tools involving external service calls that execute quickly would have negligible cost while tools executed locally would contribute higher to the cloud cost.

The number of times the tools are triggered and the type of tool also play a major role in total cloud costs.  When both the MCP server and the tools are hosted on cloud, the cloud cost would include an additional cost from the scalability of the server, which amounts to the number of request/response it handles. The architecture with the minimal cloud cost depends on the workload the MCP server is subjected to.  

\section{Future Work and Conclusions}\label{sec:fw}
We have proposed \agentx, a novel agentic workflow pattern that demonstrates competitive performance against state-of-the-art baselines in task completion, cost, and latency. Our implementation of the MCP server on a FaaS platform provides a pathway to deploy these MCP servers as a serverless function. There are several research directions that can enhance the \agentx framework and address its current limitations. By incorporating a preliminary reasoning step, similar to Chain-of-Thought (CoT), before the Stage Generation and Planner agents, the agents may produce more context-aware and logical plans, better aligned with the user's intended goal.
In its current form, \agentx lacks a dedicated recovery system. It is prone to the non deterministic nature of agentic workflow where it could potentially get stuck in a loop.

\agentx executes applications by breaking them down into stages. These stages as of now are executed in a linear fashion one after the other. However, stages that are independent of other stages can be executed in parallel, which could lead to potential latency benefits. In our current study we have only compared one of the two MCP architectures. The current system for deploying these MCP servers on cloud services is also manual. As future work, we aim to compare and contrast these MCP architectures, exploring the scalability benefits they provide, providing an automatic deployment of MCP servers on FaaS through templates to enhance the user experience, and finally extending these deployment of these MCP servers across hybrid cloud. 

The experiments can also be enhanced. The workload used currently are simple and static in nature. Testing these agentic systems on more complex workflows that involve more than three stages can help find limitations in existing systems, providing potential opportunities for improvement. The workloads themselves can further be scaled based on different attributes, e.g., number of requests being processed simultaneously or the intensity of the execution. This would also help us see the effect of execution results towards its contribution to cost reduction. These enhancements will be crucial for transitioning agentic patterns like \agentx from experimental frameworks into tools for real-world automation.

Our current study evaluates the agentic patterns only using the OpenAI GPT 4o mini LLM model. But the workflow's latency, accuracy and cost are fundamentally linked to the LLM's capabilities. A detailed evaluation across several LLM model categories is required to understand this trade-off~\cite{ipdpsw}. LLM models may also need to be run locally, for scenarios with strict privacy, cost, or latency requirements.

\section*{Acknowledgment}
This work was supported by the IBM-IISc Hybrid Cloud Lab at IISc.

\bibliographystyle{plainurl}
\bibliography{arxiv}

\end{document}